\documentclass[12pt,preprint]{aastex}

\usepackage{graphicx}
\DeclareGraphicsExtensions{.pdf,.png,.gif,.jpg,.ps}

\usepackage{psfig}
\usepackage{natbib}
\shorttitle{Observations of C8 Flares}
\shortauthors{Bowen et al.}
\usepackage{color} 
\usepackage{amsmath}           
\usepackage{amsfonts}       
\usepackage{subfigure}

\usepackage{lscape}

\def \fexv    {Fe\,{\sc xv}}
\def \fexvi   {Fe\,{\sc xvi}}

\def \fexviii {Fe\,{\sc xviii}}
\def \fexix   {Fe\,{\sc xix}}
\def \fexx    {Fe\,{\sc xx}}
\def \fexxi   {Fe\,{\sc xxi}}

\def \fexxiii {Fe\,{\sc xxiii}}

\def \s9     {S\,{\sc ix}}

\def \heii   {He\,{\sc ii}}

\newcommand{\x}{\ensuremath{\times}}

\newcommand{\ecs}{erg cm$^{-2}$ s$^{-1}$} 

\newcommand{\wm}{W m$^{-2}$}
\def\ion[#1 #2]{#1\,{\sc #2}}
\def\dens[#1]{10$^{#1}$\hskip 1.5pt{cm$^{-3}$}}
\def\densr[#1 #2]{10$^{#1}$\hskip 1pt{--}\hskip .5pt{10$^{#2}$}\hskip 1.5pt{cm$^{-3}$}}
\def\fl[#1 #2]{{#1}$\pm${#2}}
\def\orb[#1 #2]{{$#1^{#2}$}}
\def\ls[#1 #2]{{$^{#1}${#2}}}
\def\tm[#1 #2 #3]{{$^{#1}${#2}$_{#3}$}}

\singlespace
\newcounter{ion}

\begin{document}

\title{X-ray and EUV Observations of GOES C8 Solar Flare Events}
\author{Trevor A. Bowen\altaffilmark{1,2}, Paola Testa \altaffilmark{1}, 
Katharine K. Reeves\altaffilmark{1}}
\altaffiltext{1}{Harvard-Smithsonian Center for Astrophysics,
	60 Garden street, MS 58, Cambridge, MA 02138, USA; 
	tbowen@cfa.harvard.edu}
\altaffiltext{2}{Marlboro College,
	2582 South Rd, Marlboro, VT 05344, USA}

\begin{abstract}

We present an analysis of soft X-ray (SXR) and extreme-ultraviolet (EUV) observations of solar flares with an approximate C8 GOES class. Our constraint on peak GOES SXR flux allows for the investigation of correlations between various flare parameters. We show that the the duration of the decay phase of a flare is proportional to the duration of its rise phase. Additionally, we show significant correlations between the radiation emitted in the flare rise and decay phases. These results suggest that the total radiated energy of a given flare is proportional to the energy radiated during the rise phase alone. This partitioning of radiated energy between the rise and decay phases is observed in both SXR and EUV wavelengths. Though observations from the EVE show significant variation in the behavior of individual EUV spectral lines during different C8 events, this work suggests that broadband EUV emission is well constrained. Furthermore, GOES and AIA data, allow us to determine several thermal parameters (e.g. temperature, volume, density, and emission measure) for the flares within our sample. Analysis of these parameters demonstrate that, within this constrained GOES class, the longer duration solar flares are cooler events with larger volumes capable of emitting vast amounts of radiation. The shortest C8 flares are typically the hottest events, smaller in physical size, and have lower associated total energies. These relationships are directly comparable with several scaling laws and flare loop models.

\end{abstract}


\section{Introduction}
\label{s:intro}
Solar flares are transient eruptive events characterized by a localized increase in radiation from the solar atmosphere. 
Solar Dynamics Observatory (SDO) currently monitors extreme ultraviolet (EUV) emission with the EUV Variability Experiment (EVE) spectrometer \citep{Woods_2011_EVE} and the Atmospheric Imaging Assembly (AIA) narrowband imager \citep{Lemen_AIA}. The temporal cadence of these instruments proves ideal for studying eruptive solar events. In addition to enhancements in EUV intensity, flares are associated with increased X-ray radiation. The Geostationary Operational Environmental Satellite (GOES) has provided full disk measurements of soft X-ray emission (SXR) since the mid 1970s \citep{Garcia_1994}.

Many previous studies have used statistical methods to characterize and parameterize various flare behaviors: frequency distributions \citep{Drake_1971, Pearce_1988, Pearce_1993, Shimizu_1995, Feldman_1997,  Veronig_2002, Li_2012}, temporal distributions \citep{Pearce_1988}, the Neupert effect \citep{Lee_1995, Veronig_2002}, EUV and SXR enhancements \citep{Donnelly_1976, Zhang_2012, Li_2012}, energy partitioning \citep{Emslie_2012}, as well as the long term variability in flares over several solar cycles \citep{Ryan_2012_TEBBS}. These studies have utilized observations from a great variety of instruments in a variety of wavelengths (SXR, HXR, EUV, as well as H$\alpha$). Furthermore, these statistical studies of flares vary greatly in sample size, ranging from a few handfuls to tens of thousands of events. 

\cite{Feldman_1997} and \cite{Veronig_2002}  show, using GOES data, that the flare occurrence rate, as a function of peak flux, follows a power law. These studies, as well as \cite{ Drake_1971, Pearce_1988, Pearce_1993}, also demonstrate smooth distributions of flare rise, and decay, and total durations; \cite{Veronig_2002} further shows a loose but positive correlation between duration and GOES class. Similar studies have been performed by \cite{Shimizu_1995} and more recently by \cite{Li_2012}. These studies, which focus on the distribution of temporal parameters, such as duration, rise time, decay time, and flare asymmetry, typically omit analysis of non-temporal flare parameters (other than GOES class). Studies such as \cite{Emslie_2005} and \cite{Emslie_2012} analyze various energy partitions within flares but do not consider temporal aspects or thermal parameters. \cite{Feldman_1996_temperature} address the statistical behavior temperature and emission measure. Our work expands upon these previous studies through considering a variety of flare parameters.  We presently relate temporal parameters (duration, rise time, decay time, as well as intervals between various emission peaks) with thermal parameters (radiated energy, peak temperature, and emission measure), as well as spatial parameters (observed volumes and loop lengths). Previously, \cite{Aschwanden_2008} studied the statistical relationships between these quantities in both stellar and solar flares, empirically deriving several power laws for the global behavior of several flare parameters.

Unlike these previous studies, we constrain our analysis to a single GOES flare class, considering only events with an uncorrected peak observed soft X-ray flux within a small range ($\sim20\%$) around $8 \times 10^{-6}$  W m$^{-2}$. Through constraining a single quantity, such as GOES flux, we are able to relate these flare parameters with great transparency and certainty, providing for an enhanced understanding of global flare behavior. 
These C8 flares are medium sized events, occurring relatively frequently, which generally stand out strongly against background radiation. This class includes a variety of flares, ranging from the highly impulsive eruptions to long duration two ribbon events. These C8 flares may be extremely hot or relatively cool events, and may occur in small compact loops or as large filament eruptions.  Furthermore, by analyzing data from both SDO and GOES, we can determine the behavior of these flares across a broad range of wavelengths, as well as characterize relationships between X-ray and EUV radiation.  The selection of flares within a narrow GOES flux range, coupled with other selection criteria (see discussion in Section 2.4), has significantly impacted the sample size of this study, which is limited to 17 flare events occurring the period between April 2011 and August 2012.

In Section \ref{sec data_ac} we introduce the GOES, AIA, and EVE instrumentation as well as our data acquisition and processing methods; additionally, we discuss our sample of flares. Section 3 addresses the diversity of the C8 flare class as well as correlations between various flare parameters. Section 4 follows with a discussion of the trends in emissive behavior of several EUV spectral lines observed by EVE. Section 5 provides further discussion into our results, while Section 6 concludes this paper.


\section{Data Acquisition}\label{sec data_ac}

\subsection{GOES}
The GOES mission provides high cadence full disk measurements of the solar SXR flux in two channels (0.5-4 and 1-8 \AA). Flares are classified by a letter and number according to the observed peak flux in the 1-8 \AA{ }band \citep{Garcia_1994}. In this study, we only consider flares with an approximate C8 GOES class--i.e. the pre-background subtraction soft X-ray peak intensity was approximately $8 \times 10^{-6}$ \wm.

\begin{figure}[!h]
\centerline{\includegraphics[scale=0.4,angle=90]{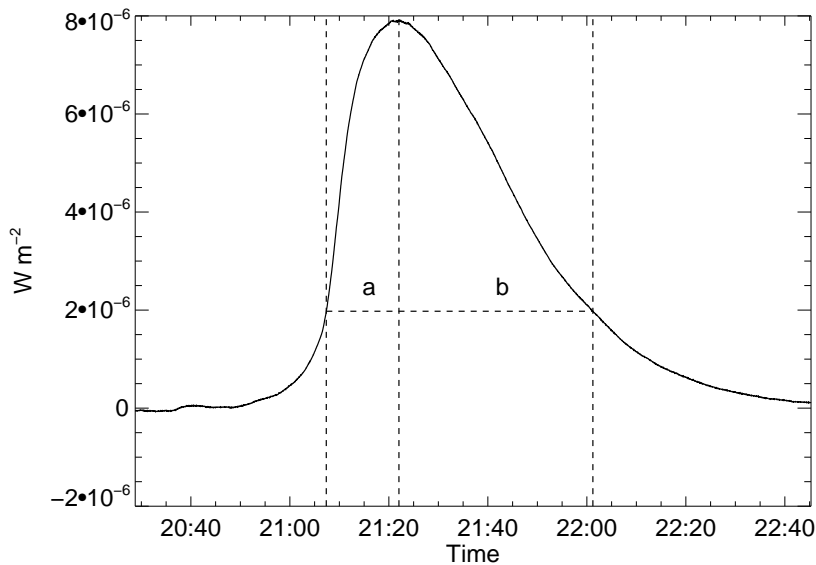}}
  \caption{Sample GOES C8 SXR flare occurring on 2011 May 29. The full width quarter-max (FWQM) duration is given by the summed rise (a) and decay (b) times.}\label{fig sample_time}
\end{figure}

We define temporal parameters of these flares using the full width quarter-max (FWQM) value of background subtracted GOES measurements (see example in Figure \ref{fig sample_time}).  The rise time, $a$, is given as the time between first quarter-max intensity and peak intensity, while the decay time, $b$, is given as the time between the peak and the return to this quarter-max value. The total duration of the flare is given as the full width quarter-max time, $a+b$. The FWQM duration includes the vast majority of the observed decay of the flare while simultaneously avoiding significant interference from the background SXR flux or other flares. 

In order to remove the contribution from quiet sun emission from our analysis, the average flux over a quiescent period preceding the flare was subtracted; a similar subtraction was used for the EVE data. \cite{Ryan_2012_TEBBS} developed an advanced method for background subtraction. Our sample is restricted to flares which are fairly isolated from other X-ray events (see Section 2.4); this constraint helps to mediate the contribution of solar background flux to the enhanced flare emission, making a more advanced background subtraction unnecessary. Table \ref{flare_table}, which lists the selected events, includes both the uncorrected flux, used in the GOES classification, as well as the background subtracted flux; the maximum deviation of our background subtracted values from the C8 flux level is approximately 20\%.
Several SSWIDL functions exist, e.g. \cite{White_2005}, which use the intensity ratio between the long and short flux channels as well as the CHIANTI atomic database \citep{Dere_1997_CHIANTI, Dere_CHIANTI_Version} to perform diagnostics of temperature and emission measure. The subtraction of the GOES background is necessary to ensure the quality of these diagnostics.

\subsection{AIA}
AIA provides imaging of the solar atmosphere with sub-arcsecond pixels on a 12 s standard cadence in a number of wavelengths \citep{Lemen_AIA}. Six of the AIA EUV channels (94 \AA, 131 \AA, 171 \AA, 193 \AA, 211 \AA, and 335 \AA) are centered on prominent spectral lines formed by various ionization states of iron. These channels contain information covering a broad range of atmospheric temperatures \citep{O'Dwyer_2010, Martinez_Sykora_2011, Reeves_Golub_2011}. The 304 \AA{ }passband is dominated by a chromospheric \heii{ }line with peak emission at Log T = 4.9. Observations from AIA are useful in studying the spatial distribution of the flaring plasma, particularly in determining flare volumes, detecting coronal mass ejections, as well as monitoring localized solar emission.  Figure \ref{fig sample_aia} displays images from the AIA 335 \AA{ }passband of six flares in our sample.

\begin{deluxetable}{ccccccccccc} 
\rotate
\footnotesize
\tablecolumns{11} 
\tablewidth{0pc} 
\tablecaption{Statistical sample of approximate GOES C8 solar flares \label{flare_table}}

\tablehead{ 
\colhead{Date}  & \colhead{Peak Time} & 
\multicolumn{2}{c}{Peak Intensity}   & \multicolumn{3}{c}{Duration} & \multicolumn{2}{c}{GOES Energy}& \multicolumn{2}{c}{MEGS-A Energy}\\ 
 & hr:min &\multicolumn{2}{c}{GOES Class}&  \multicolumn{3}{c}{min} & \multicolumn{2}{c}{$10^{28}$ ergs}& \multicolumn{2}{c}{$10^{28}$ ergs}\\
& & Observed& Corrected\,\tablenotemark{a}& Rise ($a$) &Decay ($b$) &Total ($a+b$)&Total& Rise&Total&Rise}

\startdata 
07-May-12 & 11:08 & C8.1 & C7.4 & 2.3 & 3.8 & 6.1 & 0.53 & 0.22 & 1.26 & 0.18 \\
06-Nov-11 & 09:56 & C9.1 & C8.0 & 1.6 & 7.3 & 9.0 & 0.61 & 0.17 & 5.46 & 1.38 \\
30-Dec-11 & 10:32 & C8.7 & C8.1 & 1.7 & 11.8 & 13.5 & 0.96 & 0.16 & 7.36 & 1.64 \\
15-Nov-11 & 20:34 & C7.9 & C6.7 & 6.2 & 17.9 & 24.0 & 1.66 & 0.51 & 8.11 & 1.96 \\
25-Dec-11 & 20:28 & C8.0 & C7.4 & 2.0 & 7.8 & 9.8 & 0.71 & 0.19 & 8.73 & 1.86 \\
04-Sep-11 & 01:07 & C8.4 & C7.6 & 2.3 & 15.3 & 17.6 & 1.16 & 0.22 & 10.03 & 1.63 \\
08-Aug-11 & 22:08 & C7.8 & C7.3 & 5.8 & 19.7 & 25.5 & 1.95 & 0.54 & 11.88 & 2.37 \\
21-Apr-11 & 09:47 & C8.6 & C8.0 & 5.0 & 16.9 & 21.8 & 1.89 & 0.50 & 18.97 & 4.51 \\
03-Oct-11 & 00:30 & C7.7 & C7.1 & 7.8 & 50.2 & 58.0 & 4.08 & 0.70 & 24.85 & 4.28 \\
27-Dec-11 & 04:21 & C9.0 & C8.1 & 5.8 & 19.1 & 25.0 & 1.98 & 0.58 & 25.27 & 6.66 \\
08-Jun-12 & 03:06 & C7.7 & C7.1 & 3.7 & 21.7 & 25.3 & 2.17 & 0.31 & 27.62 & 4.50 \\
30-Sep-11 & 04:00 & C7.8 & C7.3 & 9.0 & 22.7 & 31.7 & 2.44 & 0.81 & 29.34 & 8.70 \\
29-May-11 & 21:22 & C8.8 & C8.1 & 15.2 & 40.2 & 55.3 & 5.10 & 1.65 & 37.41 & 8.09 \\
09-Aug-12 & 11:47 & C8.5 & C7.5 & 9.5 & 54.8 & 64.3 & 4.09 & 0.76 & 46.32 & 6.19 \\
04-Oct-11 & 09:23 & C7.4 & C6.9 & 16.8 & 64.3 & 81.2 & 5.90 & 1.19 & 63.10 & 13.08 \\
21-Jun-11 & 03:26 & C7.8 & C7.5 & 55.3 & 99.3 & 154.7 & 12.01 & 4.78 & 88.25 & 20.23 \\
31-Aug-12 & 20:43 & C8.5 & C7.8 & 53.6 & 129.4 & 183.0 & 15.72 & 5.49 & 445.88 & 99.63 \\

\enddata
\tablenotetext{a}{GOES class (peak 1-8 \AA{ } flux) after subtraction of the background SXR emission}

\end{deluxetable}

\subsection{EVE}
EVE takes full disk spectral measurements of the EUV solar spectrum on a 10 s cadence with approximately 1 \AA{ }spectral resolution using the Multiple EUV Grating Spectrograph (MEGS) component \citep{Woods_2011_EVE}. MEGS is split into two channels: MEGS-A provides spectral data in the 65-370 \AA{ } range with a near 100\% duty cycle. This range includes lines corresponding with the seven AIA EUV channels. MEGS-B, ranging from 370-1000 \AA{ }is active for approximately three hours a day, with 5 minutes of further observations taken every successive hour. EVE observes the full disk-integrated solar spectra; accordingly, the spectral analysis of localized brightening events, such as flares, requires subtracting the pre-flare background spectrum. Typically, subtraction is performed through removing the observed spectrum averaged over a quiescent period preceding the flare. \cite{Milligan_2012_continuum} find that subtracting free-free continuum may be required for accurate spectral analysis, particularly in larger flares. The flares in our sample do not show significant continuum enhancement; accordingly, no continuum subtraction was performed. In several (three) of our events, drop out periods, lasting several minutes, occur in the EVE data set; we have used a linear interpolation to account for missing data in our light curves. Furthermore, we use a 10 min box-car smoothing algorithm in our analysis of EVE data.

The high temporal cadence, large wavelength range, and spectral resolution provided by EVE allow for the derivation of lightcurves corresponding to strong and largely unblended spectral lines which brighten during solar flares. These lightcurves provide useful diagnostics for determining the temporal evolution of thermal flare parameters.  \cite{Milligan_2012_density} used EVE observations of prominent iron spectral lines to perform density diagnostics on several GOES X-class flares. The broad range in temperatures observed by EVE, $T<10^5$ K to $T>10^7$ K, allows for an analysis of temperature evolution over the course of the flare, as shown for instance by \cite{Chamberlin_2012}, who provide a method to analyze flare heating and cooling using EVE observations.

Additionally, integrating over the spectral range of MEGS-A allows for the measurement of total broadband EUV flare emission from 65-370 \AA. These integrated EVE observations are important in comparing EUV radiation with SXR observations. \cite{Zhang_2012} show that the peak intensities of flare SXR and EUV emission are statistically correlated. Studies of the energy distribution of  flares such as \cite{Emslie_2005, Emslie_2012} and \cite{Milligan_2012_continuum} show that EUV emission corresponds to only a few percent of the total energy of a flare.

\subsection{Flare Selection}

We select flares with an approximate GOES C8 flare class. Table \ref{flare_table} lists the flares used in this study, as well as characteristics derived from GOES and EVE. Only flares for which SDO data are available were considered. These flares occurred between 2011 April 21 and 2012 August 31. Because of the pre-flare background subtraction performed on the GOES and EVE data, only flares which were adequately isolated from other events were allowed in the sample. This constraint was met by visual inspection of GOES data. Several flares during this time period were excluded because lacking EVE data could not be reasonably approximated with interpolation. Additionally, events were excluded when AIA observations suggested significant occultation of flare emission by the solar limb.

\section{Results}

\subsection{Diversity of the C8 Class}
\begin{figure}[p]
\centering
\includegraphics[width=.9\textwidth]{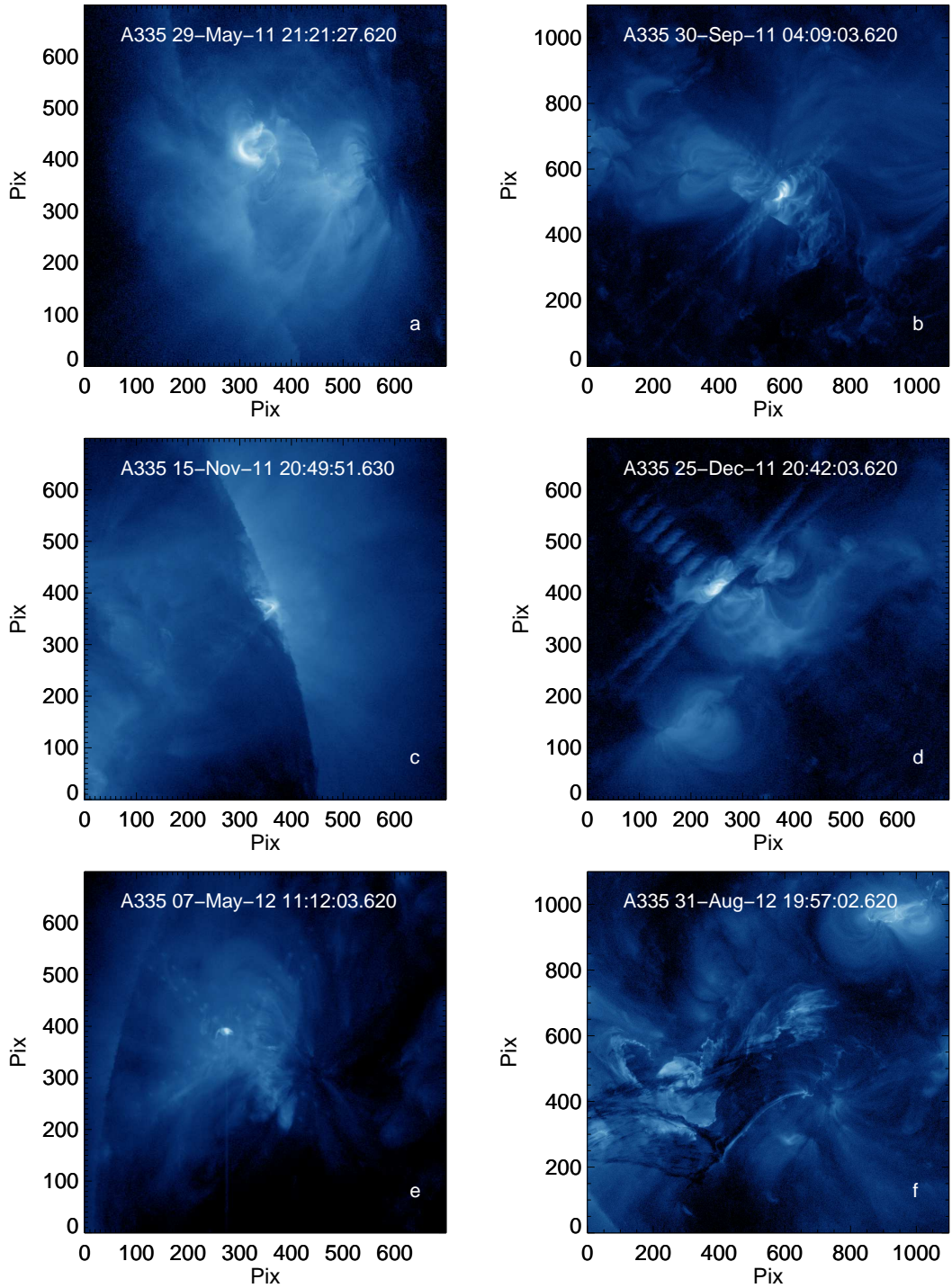}
  \caption{AIA images of six flares in our sample in the 335 \AA{ } passband. The flares displayed capture the range of events in our sample: (a) 2011 May 29. (b) 2011 September 30; (c) 2011 November 15; (d) 2011 December 25; (e) 2012 May 07; (f) 2012 August 31.}\label{fig sample_aia}
\end{figure}

\begin{landscape}
\begin{figure}[p]

\centering
\includegraphics[height=.9\textwidth]{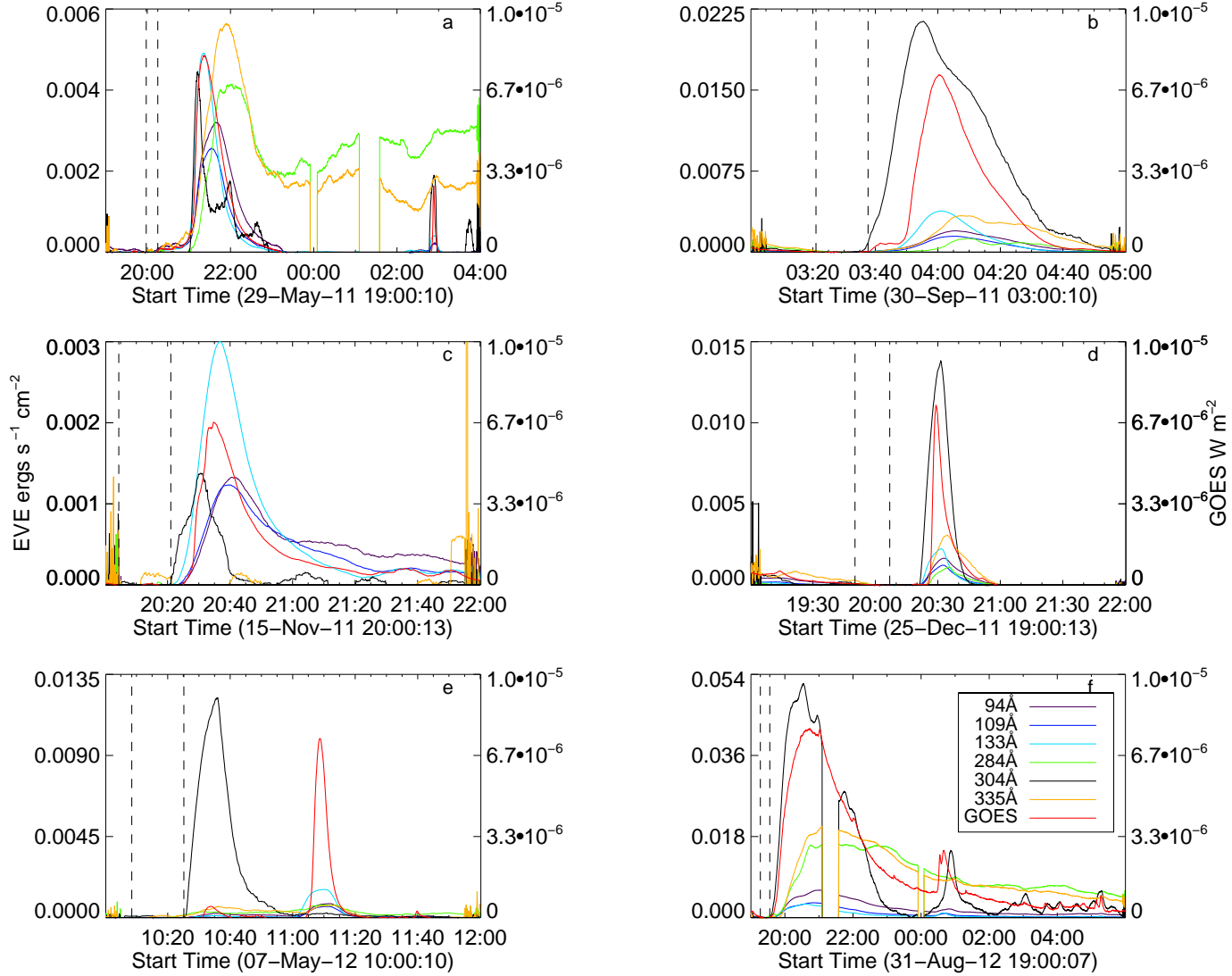}
\caption{Lightcurves of six flares of our sample. Spectral EUV lightcurves observed by EVE are displayed in \ecs\ on the left axis. The GOES SXR lightcurve (red), is given in \wm\ by the right axis. {{The dashed lines in each panel demarcate the time interval used in calculating the pre-flare background emission.}}}\label{fig sample_eve_lcs}

\end{figure}
\end{landscape}

EUV observations of these events show that the C8 flare class is highly diverse. The variation in size and morphology of these flares is evident in the AIA observations. Figure \ref{fig sample_aia}(a-f) shows AIA images in the \fexvi{ }335 \AA{ }channel of six flares included in our sample. The diversity of these flares is further exemplified by the lightcurves of individual spectral lines. EVE data were used to derive lightcurves for several prominent flare spectral lines: 94, 109, 133, 284, 304, and 335 \AA. Table \ref{tab EVE_lines} contains information regarding contributing ions as well as peak formation temperatures from CHIANTI \citep{Dere_1997_CHIANTI, Dere_CHIANTI_Version}. Figure \ref{fig sample_eve_lcs}(a-f) shows lightcurves corresponding to the six flares presented in Figure \ref{fig sample_aia}. We find that the observed emission in these spectral lines varies greatly between individual C8 flare events.

%

\begin{table}[!h]
\centering
\caption{EVE spectral lines selected for analysis}\label{tab EVE_lines}
\begin{tabular}{cccc}
\hline
\hline
Spectral Line& Peak Temp. &Temp. Range& Ions \\
\AA&Log T&Log T&\\
\hline

94 \AA&6.9&6.65-7.0&\fexviii \\ 
109 \AA&6.9&6.8-7.05&\fexix \\
133 \AA &7.1&7.0-7.3&\fexx, \fexxiii\\
284 \AA&6.3&6.25-6.5&\fexv\\
304 \AA&4.9&4.8-5.1&\heii\\
335 \AA&6.4&6.3-6.65&\fexvi\\
\hline
\end{tabular}
\end{table}


{{In the 2011 September 30, 2011 December 25, and 2012 August 31 events, pictured respectively in Figure \ref{fig sample_eve_lcs}(b, d, \& f), the 304 \AA{ }emission dominates the flare spectrum; in other events, such as 2011 November 15 and 2012 May 7 there is very little 304 \AA{ }emission, Figure \ref{fig sample_eve_lcs}(c \& e). The variability of the 304 \AA{ }\heii{ }spectral line is highlighted in Figure \ref{fig sample_eve_lcs}(e), in which the 304 \AA{ }signal from the C8 flare with a peak time $\sim$11:10 dwarfed by the emission from an earlier C1 flare which peaks at $\sim$10:30.}}

Though the 2011 May 29 flare, Figure \ref{fig sample_eve_lcs}(a), has a significant 304 \AA{ }signal, the \fexv{ }and \fexvi{ }emission observed in the 284 and 335 \AA{ }spectral lines is markedly stronger. Another particularly interesting event is the 2011 November 15 flare: flare emission is observed only in the the 304 \AA{ }as well as the hottest 94, 109, and 133 \AA{ }lines, corresponding with \fexviii, \fexix, and blended \fexx{ }and \fexxiii. This flare lacks the cooler 284 and 335 \AA{ }emission. Lightcurves of the 2011 May 29 and 2012 August 31 events show EVE data drop outs, Figure \ref{fig sample_eve_lcs}(a,f). In the subsequent analysis, linear interpolation was implemented to approximate the missing signal.
 
The flare occurring on 2012 August 31, was associated with a large filament eruption. Observations from AIA, such as Figure \ref{fig sample_aia}(f), show that this event is significantly larger and more complicated than the rest of these flares. However, the uncorrected SXR flux associated with this event only reaches a C8.5 value, meeting the criteria for inclusion in this sample. Because of the relatively small size of our sample, the magnitude and uniqueness of this event influences aspects of our statistical analysis. However, the behavior of the 2012 August 31 event is consistent with correlations and trends set by the rest of the data: excluding this event from the data does not significantly impact the results of this study. Accordingly, we do not believe it should be excluded as an outlier: though it is clearly a highly complex event, the same characteristic flare processes are in play. We note that our statistical calculations could be greatly improved by a larger sample which included more such large events.


Table \ref{tb: tab times} lists the statistical parameters characterizing the distribution of timescales within our flare sample. The mean duration of the sample is 47.4 min while the median is 25.3 min. The standard deviation, 50.6 min, is greater than the average duration, thus demonstrating a significant spread in the characteristic time scales of C8 flares.  This spread in durations is consistent with the low correlation between characteristic time and peak SXR flux found in the study \cite{Veronig_2002}, which includes flares of many GOES classes. 
\begin{table}[!h]
\begin{center}
  \caption{Characteristic FWQM flare rise ($a$), decay ($b$), and total ($a+b$) durations  }\label{tb: tab times}
\begin{tabular}{cc cc}
\hline
\hline
&$t_{tot}$ (a+b)&$t_{rise}$(a) &$t_{decay}$ (b)\\
Value & (min) & (min) &(min) \\
\hline
Mean &47.4 & 12.0&35.4\\

Median &25.3&5.8&19.7\\
Standard deviation&50.6&16.6&34.8\\
Range&6.1-183.0&1.6-55.3&3.82-129.4\\
\hline
\end{tabular}
\end{center}
\end{table}

The time scales of our sample are characteristically longer than those given in previous studies. \cite{Veronig_2002} found a mean duration for all C and M flares of $12.0 \pm 0.2$ min and $24.0\pm 1.3$ min;  rise times of $6.0 \pm 0.1$ min and $10 \pm 0.5$ min;  and decay times of  $6.0 \pm 0.1$ min and $12.0 \pm 0.7$ min. While studying the effect of X-ray background on flare occurrence rates, \cite{Feldman_1997}, analyzing a large range of classes, found that the distribution of flare durations peaks around 6-8 min; the peak of our distribution is around 15-25 min. Alternate definitions of flare start and end times contributes to the discrepancies in these time scales. \cite{Veronig_2002} use the standard GOES Solar Geophysical Data catalog start and end times; \cite{Feldman_1997} use the full width half-max duration. Reporting flare time scales using the quarter-max values can significantly extend the characteristic time intervals past half-max values. Furthermore, it is likely that our relatively small sample size, which includes the 2011 August 31 filament eruption as well as other long duration events affects the time scales which we report.

%
%
%
\subsection{Correlations in Time Scales and Energy Partitioning}\label{subsec energy_partition}
\begin{figure*}[!t]

\centerline{\includegraphics[height=\textwidth,angle=90]{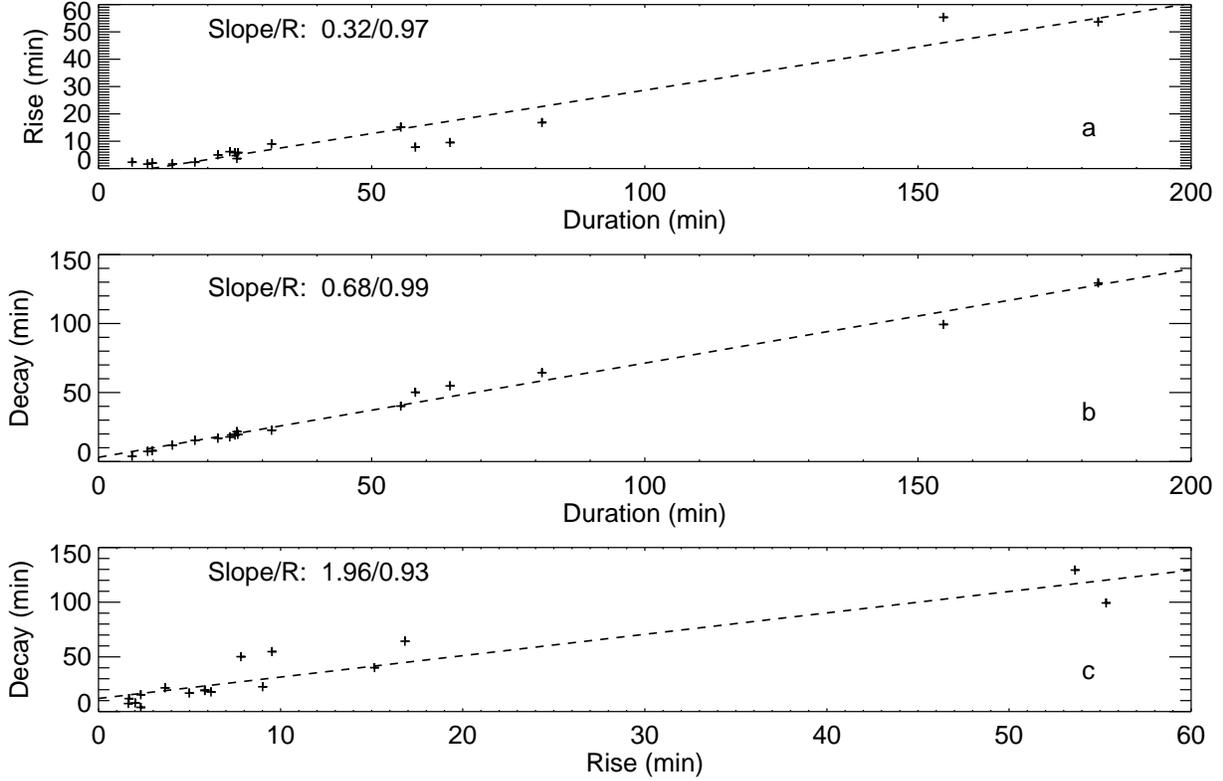}}
  \caption{A linear regression shows that the FWQM characteristic time scales of these flares are highly correlated ($R>0.90$). (a) The rise time of a flare accounts for slightly under a third of the total duration. (b) The decay from peak intensity to quarter-max intensity accounts for approximately two thirds of the duration of these flares. (c) The duration of the decay phase is proportional to the rise phase. {{In each panel, the line of best fit has been over plotted.}}}\label{fig times}
\end{figure*}

Though we find significant variability in the temporal scales of these similarly classified flares, we show that the total duration, rise, decay times of a single flare are linearly related. Figure \ref{fig times} shows substantive correlations ($R> 0.90$) between the characteristic time scales of these C8 flares. Using the FWQM definition, our statistical analysis shows that the rise phase of the flare scales as approximately a third of the total duration; while the decay phase scales as approximately two thirds the duration. Within our uncertainties, contributed to by a 10s measurement cadence, the 10 minute box car smoothing, as well as the background subtraction, a linear regression shows that the FWQM decay phase scales as twice the length of the rise phase. This linear regression gives constants of $-3.0$ and $3.0$ in relating the respective rise and decay times to the total flare duration. These constants are unphysical, it is not possible for the individual flare phases to have non-zero durations while the total flare duration is zero. Additionally, we performed this linear regression forcing this constant through the intercept; this constraint does not significantly affect the reported relationships. This proportionality between time scales has interesting implications for the partitioning of radiated energy between the different flare phases.

\begin{figure}[!h]

\centering
\includegraphics[scale=.5,angle=90]{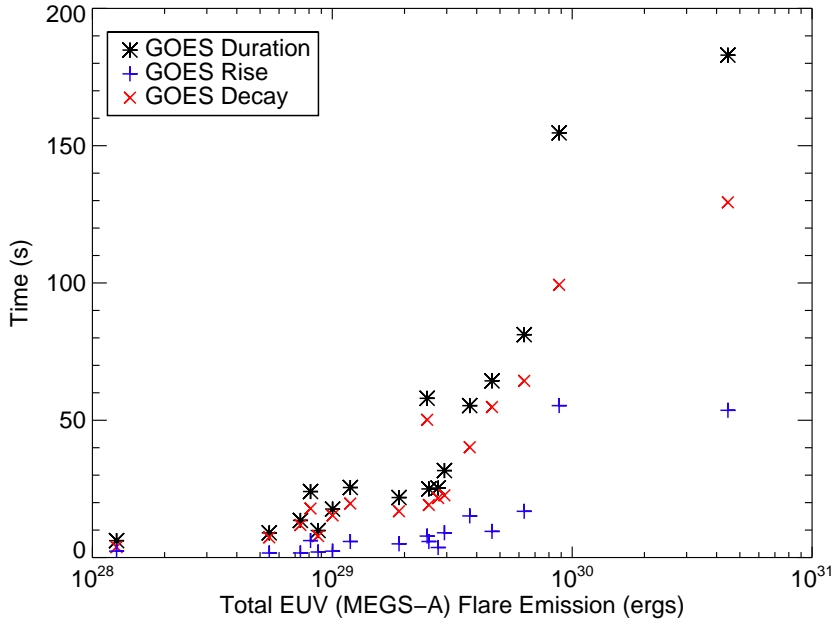}
  \caption{The characteristic FWQM times of the rise (blue + ), decay (red x) and total duration (black *) for each flare as a function of time integrated radiation in the EVE MEGS-A wavelengths.}\label{fig Time_Energy}
  \end{figure}

Figure \ref{fig Time_Energy} shows the relationship between event duration and total radiated EUV energy (broadband EVE observations integrated over the total flare duration). Energy is positively correlated with flare duration: this is not surprising as the total energy is taken as an integral over time.  Figure \ref{fig Time_Energy} also shows the X-ray energy portioned into the flare rise and decay phases. We find that the decay phase always contributes more to the total energy than the rise phase.

\begin{figure}[!h]
  \centering
  \includegraphics[scale=0.5,angle=90]{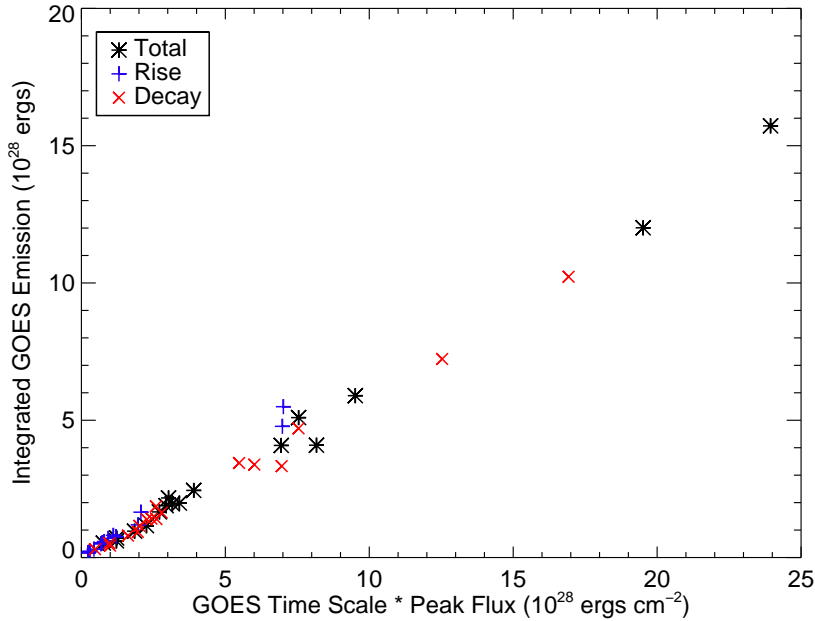}
    \caption{ Time integrated GOES emission over the total flare, as well as the decay and rise phases plotted versus the duration of the appropriate time interval multiplied by the peak GOES flux.}\label{fig Veronig_compare}
\end{figure}

 Figure \ref{fig Veronig_compare} shows a plot of the total radiated SXR emission as a function of duration multiplied by the peak flux. We find that these two quantities are highly correlated. This is consistent with the findings of \cite{Veronig_2002} and \cite{Li_2012} who each looked at a range of flare classes. \cite{Veronig_2002} interprets this result as an indication that the profile of the GOES lightcurve is not strictly important to the time integrated flux of a flare.
 
Additionally, Figure \ref{fig Veronig_compare} demonstrate that the emitted energies in the individual rise and decay phases follow this relationship: the amount of energy radiated in the rise and decay phases is directly proportional to the peak flux multiplied by the duration of each phase. Because our sample is constrained  to flares of a similar peak SXR flux, it follows that radiated energy, in either the total, rise, or decay phase these flares, is simply proportional to the relevant time scale.

\begin{deluxetable}{ccccccccccc} 

\footnotesize
\tablecolumns{11} 
\tablewidth{0pc} 
\tablecaption{Impact of time definitions on  proportionality constant between rise and main phase of the flares in our sample and associated correlation coefficient ($R$).\label{time_defs}}

\tablehead{ 
\colhead{Definition} & \multicolumn{2}{c}{Time}  & \multicolumn{4}{c}{GOES Energy} & \multicolumn{4}{c}{MEGS-A Energy}\\

&Slope &$R_{time}$&Power&$R_{log}$&Slope &$R_{lin}$&Power&$R_{log}$&Slope &$R_{lin}$}

\startdata 
FWHM&0.37 & 0.97 & 0.98 & 0.95 & 0.40 & 0.98 & 1.00 & 0.98 & 0.32 & 1.00 \\
FWQM&0.32 & 0.97 & 1.03 & 0.95 & 0.37 & 0.98 & 1.03 & 0.99 & 0.25 & 1.00 \\
FW1/8M&0.26 & 0.96 & 1.07 & 0.95 & 0.34 & 0.98 & 1.04 & 0.99 & 0.19 & 1.00 \\
FW1/16M&0.18 & 0.91 & 1.06 & 0.95 & 0.30 & 0.97 & 1.00 & 0.98 & 0.14 & 1.00 \\

\enddata

\end{deluxetable}

\begin{figure}[h]
\includegraphics[height=\linewidth,angle=90]{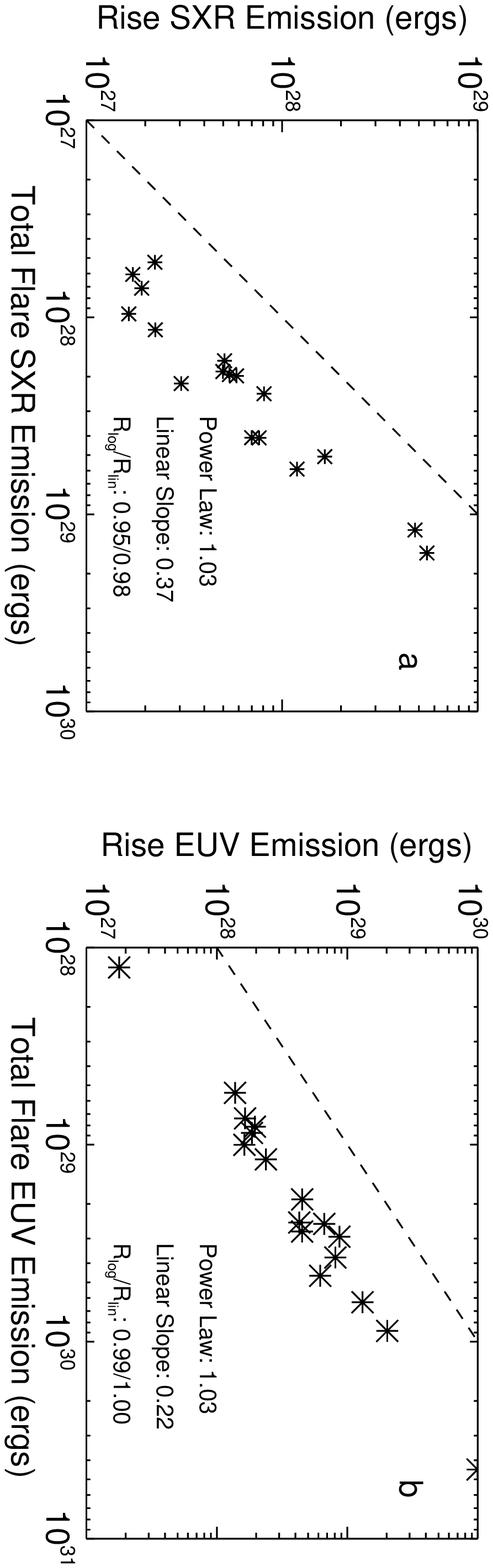}

\caption{For both GOES (a) and broadband EVE (b) observations, a regression in log space shows that the energy emitted over the rise phase is linearly related (power law with index 1.03 in either case) to the total integrated emission. The dashed lines in each plot depict a perfectly linear slope. An additional linear regression gives the constant of proportionality between the rise and total energies: 0.37 for GOES and 0.22 for EVE.}\label{fig total_rise}
\end{figure}

We find high correlations between the total duration of a flare with the individual durations of its rise and decay phases. Additionally, we have shown that the radiation emitted over each phase of the flare is simply proportional to the duration of the respective phase. Together, these results suggest a constant partitioning of energy between the different phases of a flare. Figure \ref{fig total_rise} shows that the emission observed in the GOES 1-8 \AA{ }waveband over the total flare duration is closely correlated to the energy radiated in the rise phase alone. A regression in logarithmic space shows that the energy radiated in the rise phase of a flare scales linearly with total SXR emitted energy (a power law index of 1.03 with $R=0.95$). A regression in linear space gives a line of best fit with a slope of 0.37 and a strong correlation value of $R=0.98$. {{We interpret this result as an indication that C8 flares radiate slightly under 40\% of their total 1-8 \AA{ } X-ray emission during their rise phase.}}

%


Integrating over the EVE wavelength range provides a method to analyze broadband flare EUV radiation. Figure \ref{fig total_rise} also shows high correlations in the time integrated EUV radiation between the rise phase and total duration. Emission observed by EVE during the GOES FWQM rise phase scales linearly (power law index of 1.03) with the emission observed over the total flare FWQM duration. {{A best fit line gives a constant of proportionality of 0.22, with a correlation of $R>0.99$. Accordingly, approximately 22\% of the broadband EUV emission of a typical C8 event occurs during the rise phase.}} The linear constant from this fit is close to $2.5 \times 10^{27}$ ergs. 
We note that these constants are unphysical quantities used only to define our line of best fit; in the limit of zero total radiated emission, the rise energy should clearly tend to zero.

It is striking that a linear correlation between energies emitted in the rise and decay phase of the flares occurs in both EVE and GOES observations. The difference in percentage of energy partitioned in the rise phase between these two instruments {{{($\sim37$\% for GOES and $\sim22\%$ for EVE) follows from differences in their temperature response functions. GOES is sensitive to high temperature plasmas present early in the flare, while broadband EVE observations include many cooler spectral lines. Emission from these EUV spectral lines, brightest primarily in the decay phase of a flare, are a significant portion of the total EUV radiation. Accordingly, compared with GOES X-ray observations, the rise phase EUV emission accounts for a lower percentage of the total EUV flare radiation.}}}

The correlations found between characteristic flare times, as well as  SXR and EUV energies, are sensitive to the definition of the FWQM time scale. We further investigate the relationship between rise and decay times using several alternate definitions: full width half-max (FWHM), full width eigth-max (FW1/8M), and full width sixteenth-max (FW1/16M). In considering these factors, we consistently find highly correlated linear relationships between the flare time scales. Table \ref{time_defs} shows the various linear slopes and correlation values calculated for the time, SXR, and EUV energies. 

The use of FW1/8M and FW1/16M values causes the slope relating the duration of the rise phase relative to the total flare duration to decrease. Because of the impulsive nature of the rise phase, alternate definitions of the start time do not greatly affect the length of the rise phase. However, the decay time, and thus the total duration, is significantly affected by these distinctions. 
Though the use of alternate time definitions impacts the reported durations of these events, we continue to find linear relationships between rise time and total time.

Similarly, using alternate definitions primarily affects the energy reported in the flare decay phase. However, Table \ref{time_defs} shows that a highly correlated energy partition exists regardless of particular time definitions used. In exploring time definitions which extend the flare duration (i.e. the FW1/8M and FW1/16M), we find that the proportion of flare energy partitioned in the rise phase decreases. 

\subsection{Additional Flare Parameters: Temperature, Emission Measure, Volume}\label{subsec TEMP}
\begin{figure}[t]
\centerline{\includegraphics[scale=0.4,angle=90]{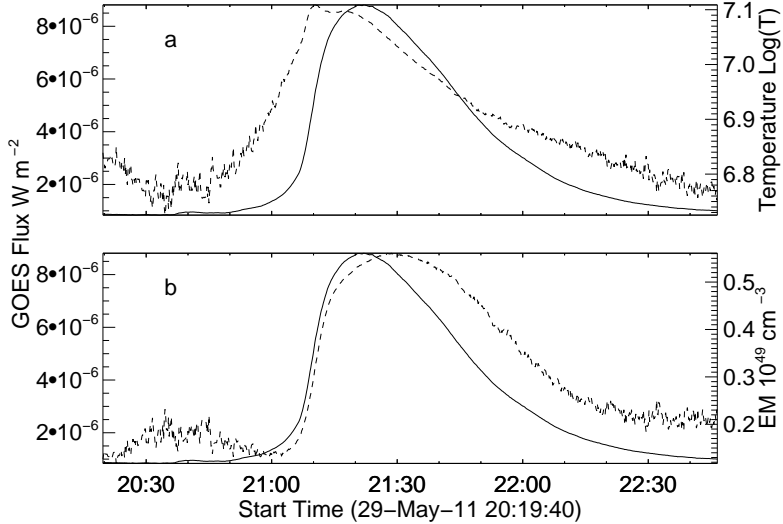}}
  \caption{Dashed line shows a sample diagnostic of temperature (a) and emission measure (b) from the 2011 May 29 flare. The 1-8 \AA{ }GOES flux is plotted as a solid line in both frames.}\label{fig sample_T_EM}
\end{figure}

In addition to time scales and broadband radiated energy, we use our constraint of GOES class to consider a range of flare properties. In particular, we address the thermal parameters for the flaring plasma, such as temperature, emission measure, and volume. Temperature and emission measure have been obtained through diagnostics of the GOES lightcurves, while volumes are estimated from the AIA images.

 Observed GOES flux is given by the expression \begin{equation}F_{SXR}=G_{\lambda}(T)N_e^2V,\end{equation} where $G_{\lambda}(T)$ represents temperature response of the GOES channels, and $N_e$ and $V$ are the density and volume of the flare plasma. The emission measure is given by the term $EM=N_e^2V$. Because $G_{\lambda}(T)$ is well known for both GOES channels, the ratio between the long and short channel observations allows for a diagnostic of flare temperature and emission measure. Several SSWIDL functions exist allowing for the straightforward calculation of these parameters under an isothermal assumption \citep{White_2005}. Though \cite{Reeves_Warren_2002, Warren_2006, Reeves_2007} have shown that flares likely consist of many multi-thermal strands, the approximation provided by the isothermal assumption suffices to demonstrate trends in temperature and emission measure between these flares.

\begin{figure}


\includegraphics[scale=.55,angle=90]{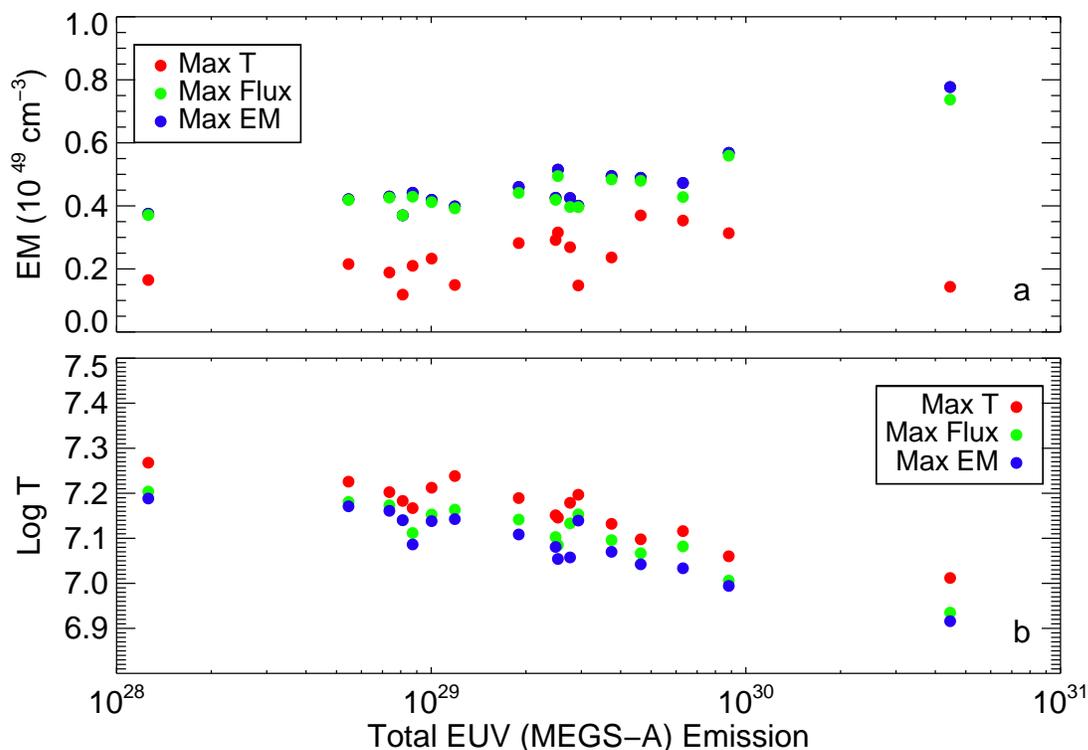}
\vspace{-.3in}{{
  \caption{{{(a) Flare emission measures at time of maximum temperature (red), peak SXR flux (green),  and maximum emission measure (blue) as a function of the EUV radiated energy.  Within our sample there is little deviation in the emission measure at the time of max temperature, which occurs early in the flare. The emission measure at peak flux (green) is typically similar to maximum emission measure (blue); these values are usually a factor of 2-3 larger than the initial values. (b) Logarithmic temperature at time of maximum temperature (red), peak SXR flux (green), and maximum emission measure (blue). The hottest events are generally less emissive flares. This figure additionally helps convey the time evolution of flare temperature and emission measure; chronologically, max temperature (red) occurs first, followed by peak SXR flux (green), and finally maximum emission measure (blue).}}}\label{fig temp emis}}}
\vspace{.1in}
\end{figure}

Figure \ref{fig sample_T_EM} shows the characteristic behavior of temperature and emission measure for one flare of our sample. Maximum temperature is attained early in the lifetime of the flare, typically close to the quarter-max onset. GOES X-ray flux peaks later in the event, either preceding or occurring simultaneously with the maximum emission measure. {{Figure \ref{fig temp emis}(a \& b) shows flare temperature and emission measures at the time of maximum temperature, peak GOES flux, as well as maximum emission measure. We demonstrate strong scaling with these parameters and the total radiated EUV energy. It is important to note that, because of our constraint on peak GOES flux, temperature and emission measure are not both free parameters (see Equation 1). Accordingly, the scaling shown between, for instance, flare temperature with the total radiated energy determines the scaling between emission measure and energy. This figure shows that the more energetic, accordingly longer duration, events of our sample of C8 flares are cooler;  in order for cooler events to have a peak C8 flux, these flares must have characteristically higher emission measures. Additionally, this figure highlights the large range of temperatures and emission measures which can be attained by C8 events.}}

\begin{figure}[p]
\centering
\includegraphics[width=\textwidth]{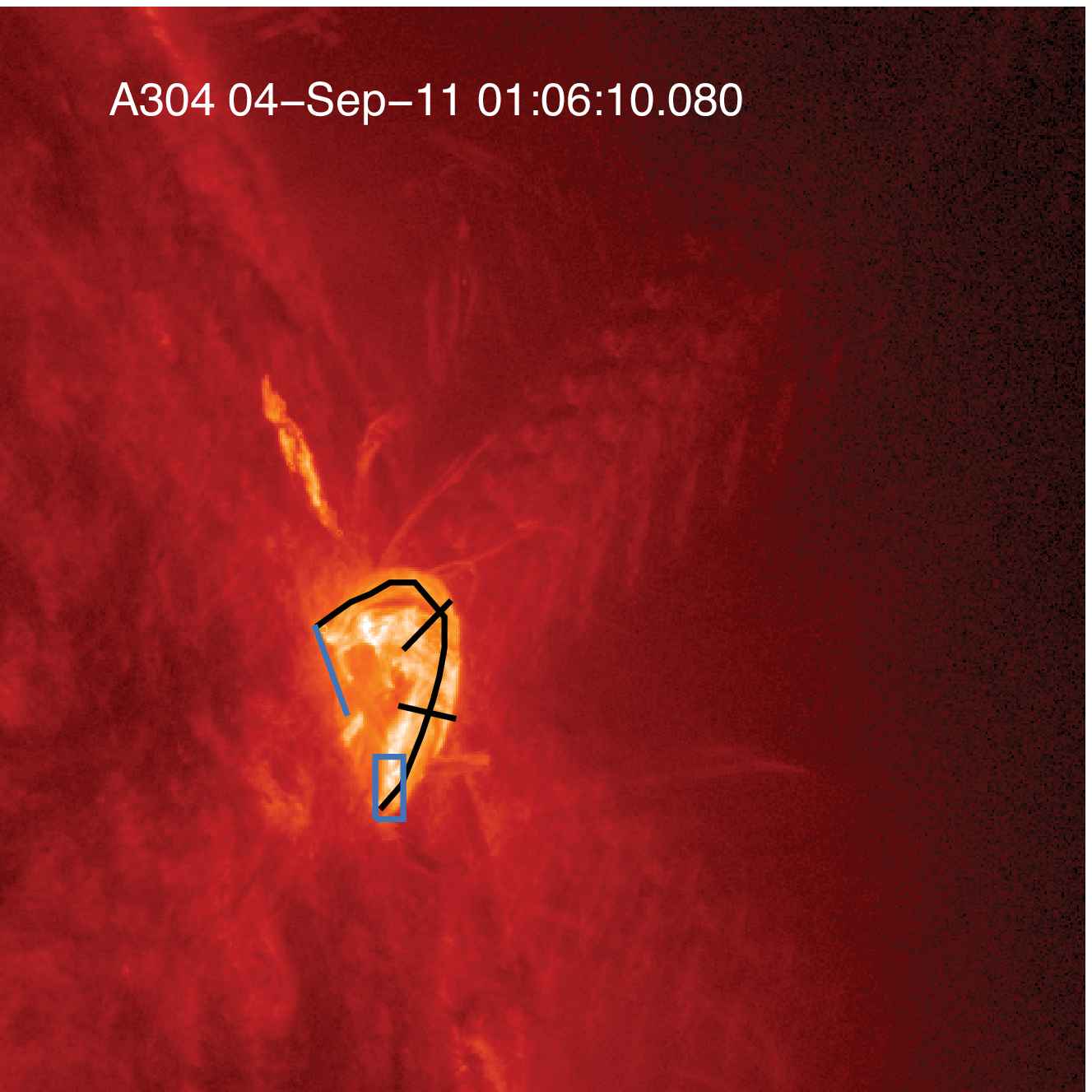}
  \caption{AIA 304 \AA{ } image of the 2011 September 04 flare showing our methods for estimating flare volume. The (black) lines plotted over the flare loop show estimates for flare length, depth, and width. The (blue) radial line and box shows our estimated loop radius and footpoint area. The volume was taken as the average of these two estimates. This flare has a clear loop like structure; other flares, particularly those positioned further from the limb, lack such an easily identifiable loop structure.}\label{fig sample_dimens}
\end{figure}

{{
Analysis of AIA data allows for the estimation of flare volume using the average of two manual calculations. One method approximates the flare arcade as a semi-circular structure with a cross-sectional area defined by the footpoints. Using AIA image data, we manually select a radius $r$ to fit a semi-circular structure to the flare arcade.  The footpoint area, $A$, is determined by drawing a box that encompasses the flare loop footpoints in the image.  The volume of the flare is given by the circumference of the semi-circle times the footpoint area ($\pi r A$). The other method multiplies estimates of the depth, width, and length of the flare arcade structure. Figure 10, which shows an AIA 304 \AA{} image taken of the 2011 September 04 flare, shows measurements of the flare volume using both methods. These volume estimations were performed with the AIA data which provide clearly resolvable and localized flare emission representative of the total extent of the flare. }}{{The uncertainties on these measurements, arising primarily from projection effects, can be quite large. In some cases, flare morphology, as well as orientation and location on the solar disk, prohibit precise measurements of flare geometry; in these analyzing these events, uncertainties in volume measurements may be close to an order of magnitude. Other events exhibit  clear geometry which are relatively easy to measure (e.g Figure \ref{fig sample_dimens}).Because of the large systematic error arising from projection effects, as well as the inherent subjective in manual measurements, a strict quantification of these uncertainties proves quite challenging.}}

\begin{figure}[p]
\centering
\includegraphics[scale=.6,angle=90]{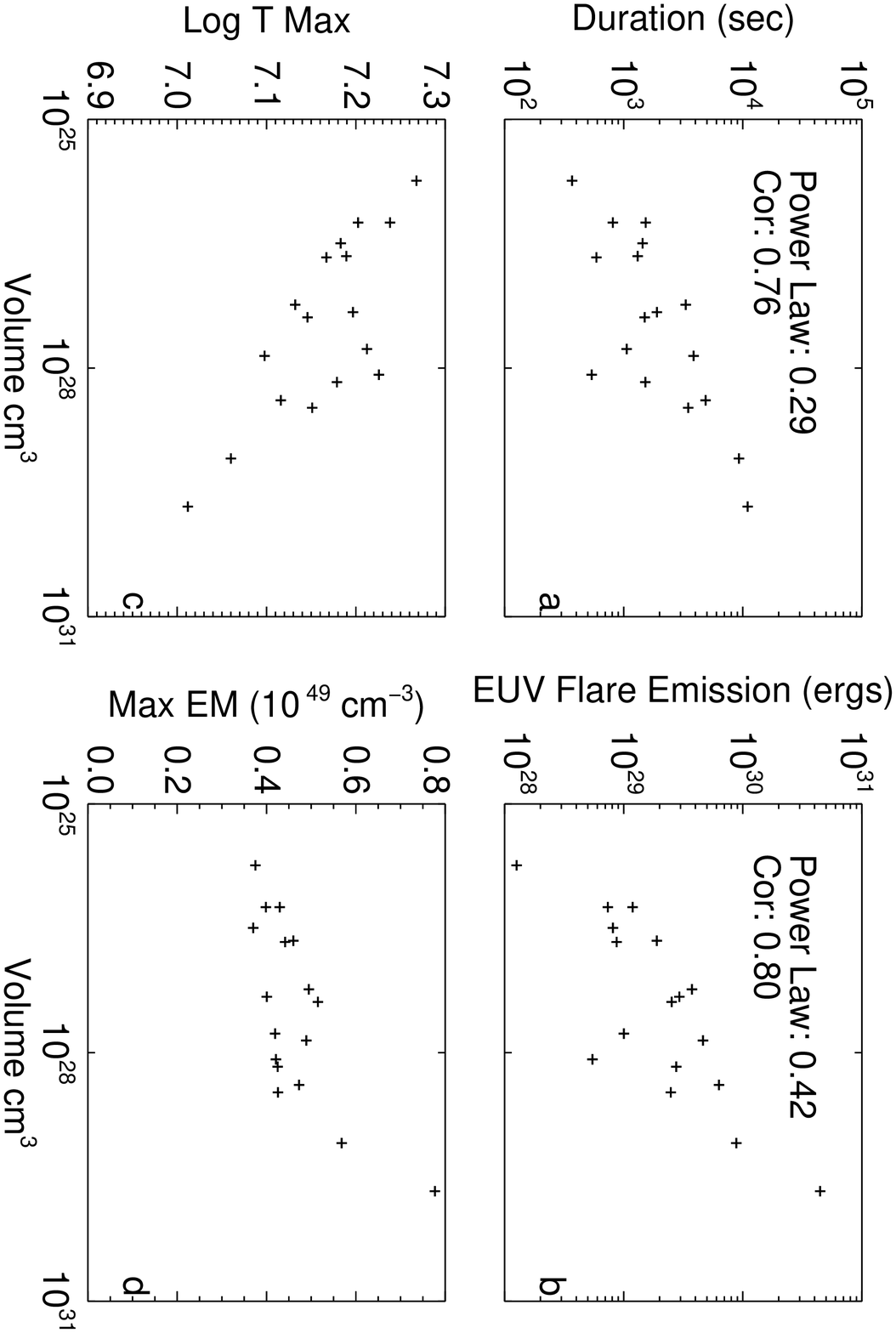}
\vspace{.4in}
  \caption{(a) Flare duration as a function of volume; there is a clear positive correlation. (b) Flare emission as a function of volume, once again there is a significant positive correlation. (c) The logarithmic temperature is negatively correlated to flare volume. (d) The emission measure seems to have a small positive correlation with volume; we note a decent scatter, more data points are needed to fully understand this trend\label{fig Vols}.}

\end{figure}

Volumes of these flares vary from approximately $5\times10^{25}$ cm$^3$ to $5\times10^{29}$ cm$^3$. Figure \ref{fig Vols} shows relationships between volume and other various flare parameters. There is a significant relation between flare volume and duration. We find a power law slope of $a=0.29 $ and a correlation of $R=0.76$. The longer flares of this sample tend to involve larger volumes of flaring plasma, and larger total emitted energies (as shown in the previous section). We find a power law value relating volume and total emission of $a=0.42$ and a correlation of $R=0.80$. {{Additionally, we find that the larger volume C8 flares are generally cooler events; in these flares the input energy is spread over a much greater volume of plasma.}}

%
%
\begin{figure}[t]
\includegraphics[scale=.6,angle=90]{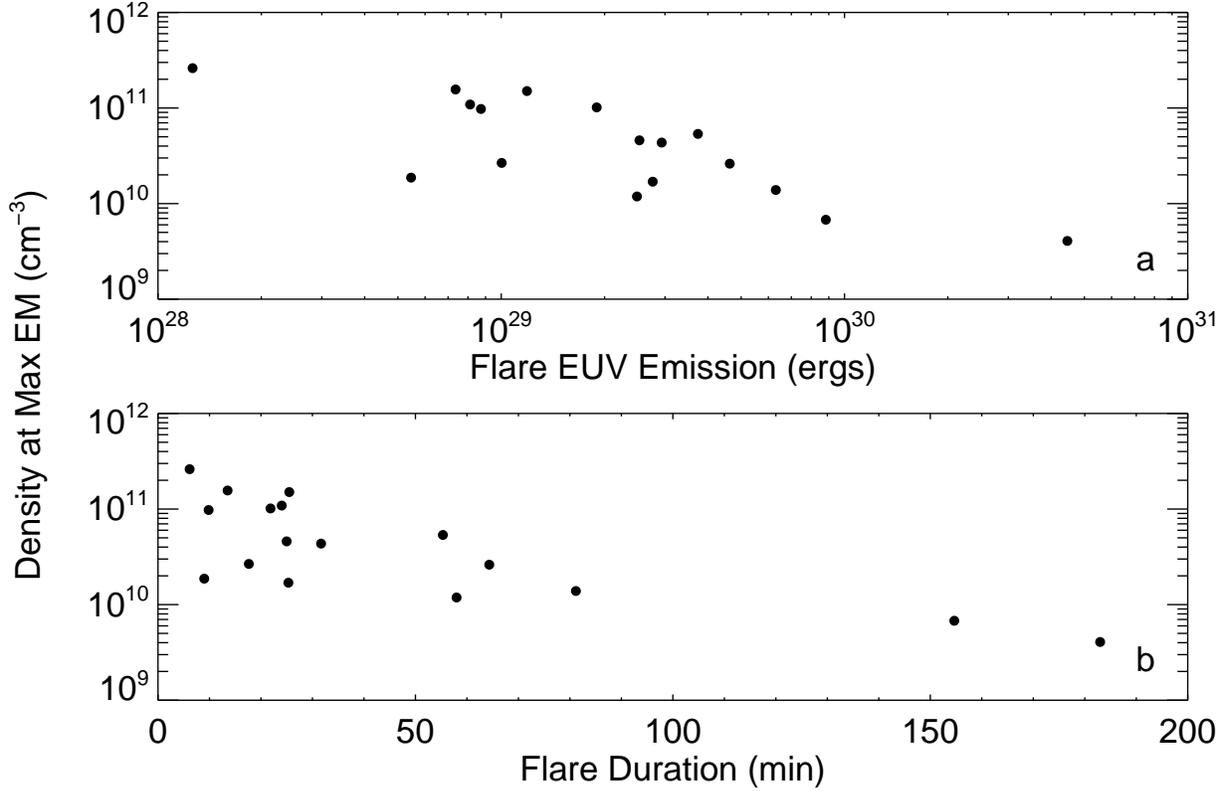}
  \caption{(a) Density at the time of maximum emission measure as a function of total radiated EUV energy. (b) Density at time of maximum emission measure as a function of FWQM GOES duration. It is clear that the longer and more emissive flares have lower densities}\label{fig dens}
\end{figure}

Volume estimates, together with emission measure values diagnosed from the GOES filter ratio analysis, provides an estimate for the density of the flaring plasma. Assuming constant volumes, the density is simply proportional to the square root of the emission measure:
$$N_e=(EM/V)^{1/2}.$$
The densities at the maximum emission measure are shown in Figure \ref{fig dens} as functions of both the flare duration and radiated EUV energy. The highly emissive flares with the longer durations and greater volumes tend to have lower densities. It is important to note that these values represent the density at the time of maximum emission measure; the highest densities $10^{11}-10^{12}$ cm$^{-3}$ occur in the flare footpoints during the impulsive phase \citep{Milligan_2011}. \cite{Milligan_2012_density} provide an alternative method for density diagnostics through the use of several \fexxi{ }lines without knowledge of the flare volume.

\begin{figure}[!t]
\centering
\includegraphics[scale=.6,angle=90]{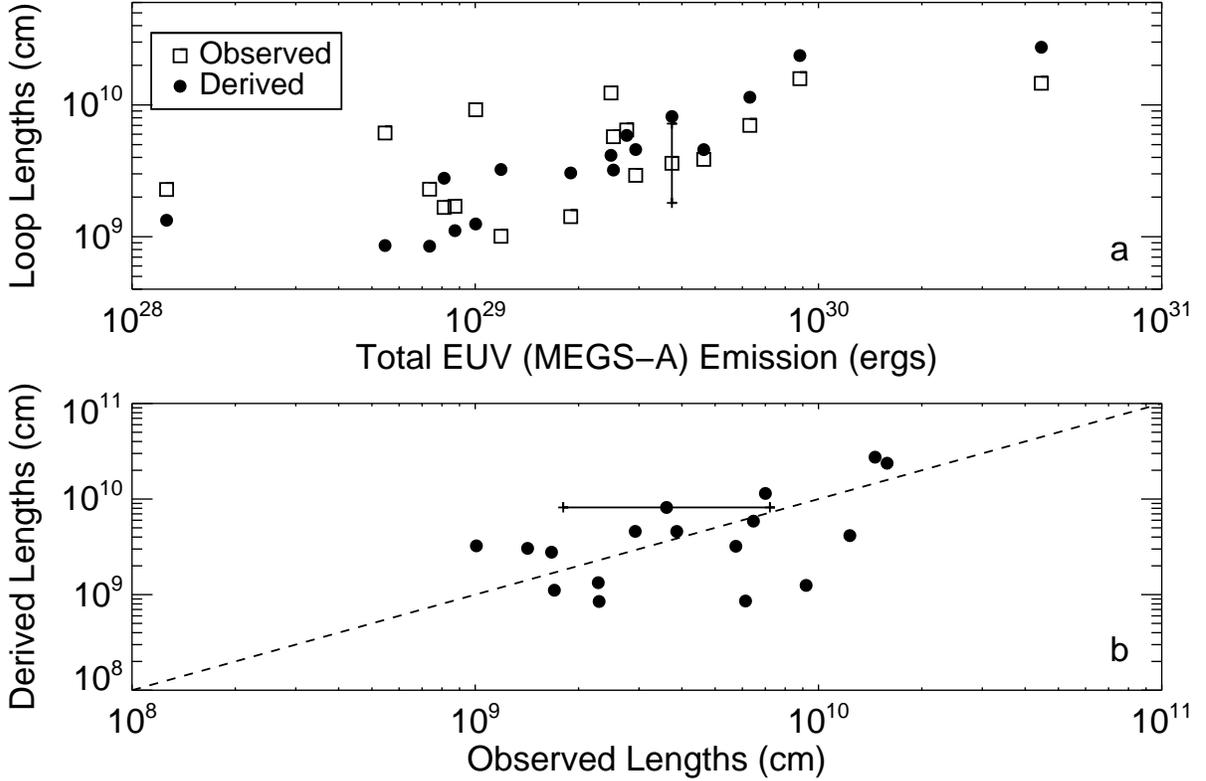}
  \caption{(a) Observed and derived  flare loop lengths. Observed lengths were taken from manual measurements of AIA observations while derived lengths are given by the method of \cite{Reale_2007}. (b) Derived flare loop lengths plotted against observed lengths; the dashed line shows a 1:1 correspondence. {{ Error bars have been plotted on one data point to illustrate the factor of two uncertainty on the observed loop lengths.}}}\label{fig loop_lens}
\end{figure}

The results of our analysis are useful in testing flare models and scaling laws. For example, \cite{Reale_2007}, using the previous work of \cite{Serio_1991}, proposes hydrodynamic scaling laws for the cooling times and lengths of a single flaring loop. Diagnostics of flare temperature and emission measure from the GOES channels allow for a comparison of observations with this simple loop model, which gives an expression for loop length as  $$L=3\x10^{3}\left(\frac{T_0}{T_M}\right)^2 T^{1/2}_{0,7}t,$$ {{where $L$ is the half loop length in units of $10 ^9$ cm, $T_0$ is the maximum temperature, $T_M$ is the temperature at the time of maximum density (i.e. emission measure), $T_{0,7}$ is the  peak temperature of the plasma in units of $10^7$ K, and $t$ is the time of the maximum density in seconds.}} Figure \ref{fig loop_lens} shows the diagnosed loop lengths with the estimated lengths taken from AIA data. {{Though highly quantifiable uncertainties are difficult to determine for these loop lengths we are highly confident that these estimations are accurate within a factor of two. The derived lengths are, for the most part, within this range of uncertainty. The dashed line in the bottom plot shows the 1-1 correspondence; we note that the data roughly follow this trend, but with a notable scatter. }}

%


\section{EVE lightcurves}

We have established that these flares vary significantly in their energetic, temporal, and thermal aspects. Understanding the correlations between the temporal and thermal parameters allows for deeper analysis of the variability in EUV spectral emission. For instance, we show how the lower temperatures and densities of the longer duration C8 events help explain aspects of the diverse spectral signatures observed by EVE.

\begin{figure}[!h]
\centerline{\includegraphics[scale=.4,angle=90]{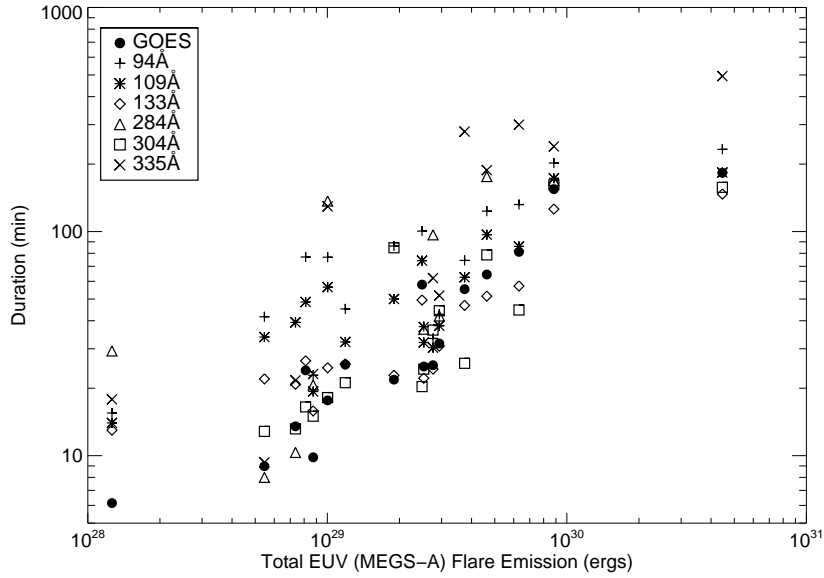}}
\caption{FWQM durations of EVE spectral lines lightcurves; GOES SXR duration is also displayed.}\label{fig lc_durations}
\end{figure}

Durations for each lightcurve were calculated with the FWQM method used in determining characteristic times from GOES. Figure \ref{fig lc_durations} shows the duration of each lightcurve, including the GOES 1-8 \AA{ }channel, as a function of the total radiated EUV energy.  Each spectral line shows the expected correlation between duration and total radiated EUV output: longer duration flares are more emissive. The 335 \AA{ }duration shows a particularly high dependence on the total emitted radiation; for the largest events, this \fexvi{ }emission lasts significantly longer than the hotter spectral lines. In both the impulsive and intermediate flares, 335 \AA{ }emission endures on time scales similar to the hotter spectral lines.

For several flares in the sample, the cooler \fexv{ }and \fexvi{ }spectral lines do not return to pre-flare levels--e.g. 284 \AA{ }emission from the 2011 May 29 event displayed in Figure \ref{fig sample_eve_lcs}. This effect occurs in three 284 \AA{ }and four 335 \AA{ }lightcurves. The lack of full decay is typically attributed to either interruptions by other events or the presence of late phase emission outlined by \cite{Woods_2011_latephase} as well as \cite{Hock_2012}. Because these lightcurves do not have a determinable FWQM duration, they were omitted from Figure \ref{fig lc_durations}. Though observed in several of the flares in our sample, full consideration of late phase emission is out of the immediate scope of this study. Other data were omitted when flare emission in a given wavelength did not significantly increase above background levels: e.g., the cooler spectral lines in the 2011 November 15 event.

\begin{figure}[!h]
 
 \centering
\includegraphics[scale=.4,angle=90]{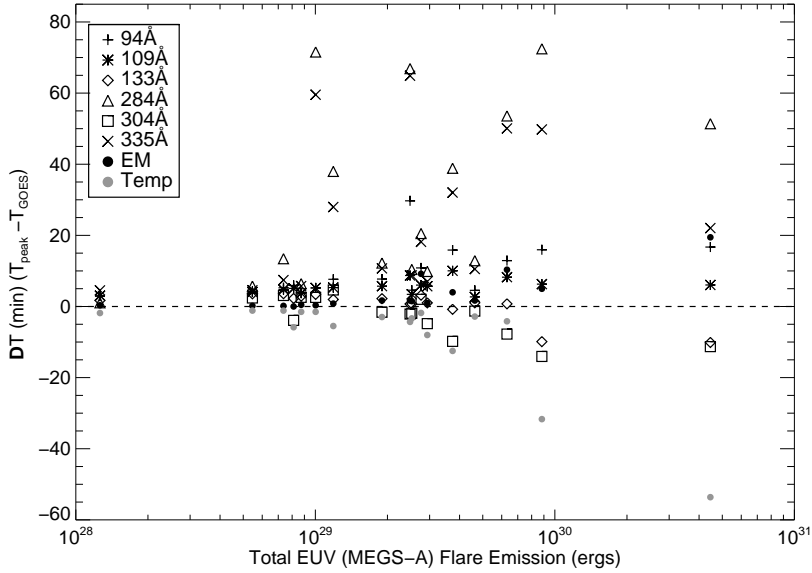}
\caption{Peak times of EVE spectral line relative to the time of the peak GOES 1-8 \AA{ }emission: zero corresponds to the time of the GOES peak. The times of peak temperature and emission measure are also shown.}\label{fig lc_peaks}
\end{figure}

Flare cooling can be demonstrated through analysis of the peak emission times of different spectral lines. Each lightcurve (except 304 \AA) is dominated by emission from a specific ionization state of iron. These ionization states have a broad range in formation temperature. Accordingly, the time elapsed between each lightcurve peak is indicative of the thermal evolution of the flare plasma.  Figure \ref{fig lc_peaks} displays the intervals between the peak time of each lightcurve and the time of peak GOES flux, negative values mean the lightcurve attained a maximum value before the GOES lightcurve; once again this figure omits non-flaring spectral lines. \cite{Chamberlin_2012} use a similar technique to show the thermal evolution of several flares through EVE data. 

 In the hotter and less energetic events, each lightcurve peaks within a few minutes of the others, indicating the rapid evolution in these flares. The intermediate events show a similar behavior in the hotter spectral lines; in contrast, these events show a significant scatter in the the peak times of the cooler 284 and 335 \AA{ }lightcurves. As expected, we find that 335 \AA{ }emission consistently peaks before the slightly cooler 284 \AA. Time intervals between successive lightcurve peaks are increased in the flares which emitted the most radiation. The \heii{ }304 \AA{ }line, with a peak formation temperature around 80,000 K, is indicative of impulsive chromospheric heating. For low and intermediate energy flares, maximum 304 \AA{ }emission occurs within the several minutes around the GOES peak; alternatively, with the longer duration flares, 304 \AA{ }emission is a consistent precursor to the peak GOES flux.

The timing of peak temperature and emission measure with respect to the peak intensities of the GOES and EVE lightcurves, also shown in Figure \ref{fig lc_peaks}, can help explain phenomena observed in the EVE lightcurves. Spectral line intensity, given by the equation $I=G_{\lambda}(T)N_e^2V,$ is highly dependent on flare temperature and emission measure. The evolution of these parameters are key in the timing and intensity of the EVE lightcurves. 
 
Though the peak intensities of 133 \AA{ }and SXR emission generally occur at the same time, 133 \AA{ } emission peaks significantly earlier in the two most energetic flares. In these same events, the maximum temperature also occurs significantly earlier relative to the other events in our sample. The 133 \AA{ }spectral line and GOES are both sensitive to temperatures greater than Log T=7.1; however, the GOES 1-8 \AA{ }response function also has significant contribution from lower temperature plasma. The advancement of 133 \AA{ }emission occurs because of the relative early timing of the peak temperature. In contrast, the maximum GOES flux, can occur when peak when temperatures are lower.

\begin{figure}[!h]
 \centering
\includegraphics[scale=.4,angle=90]{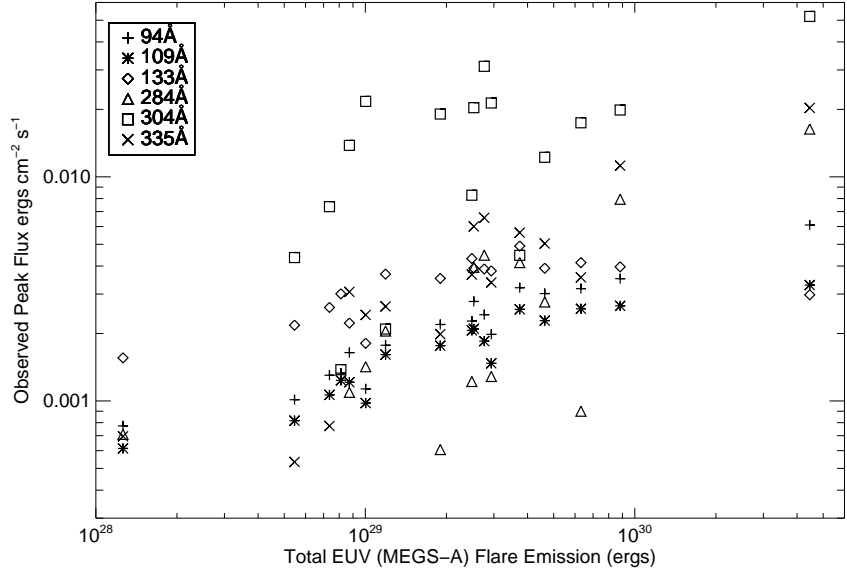}
\caption{Max intensities of EVE spectral line lightcurves.}\label{fig max_lc_intens}
\end{figure}

Additionally, the later occurrence of peak emission measure suggests that cooler spectral lines should be more emissive in longer duration events: peak emission measure occurs when the plasma temperature is lower. Figure \ref {fig max_lc_intens}, which shows the maximum intensities achieved by each lightcurve as a function of the total EUV energy, is consistent with this hypothesis. Other than the 304 \AA{ }emission, which is the result of particle deposition and heating of the chromosphere through conductive fronts, the cooler 284 and 335 \AA{ }spectral lines are notably bright in long duration events, dominating over the hotter spectral emission. Figure \ref {fig max_lc_intens} furthermore shows that the peak flux in the 94 and 109 \AA{ }wavelengths have upward trends with the total emitted EUV energy. This result is consistent with our finding that the emission measure remains enhanced though the peak temperature has decreased; not only do lines formed under Log T = 7 peak later in the most energetic events, they are also more emissive. This upward trend is also contributed to by the slight positive correlation in the size of emission measure with flare duration and energy reported in Section 3.2.  

In comparison with the other spectral lines, the peak intensity of the 133 \AA{ }emission is fairly constant across our sample. The formation temperature of the 133 \AA{ }is similar to the GOES 1-8 \AA{ }X-ray emission. Because the flares in our sample have a constrained peak GOES flux, it is not surprising that the magnitude of 133 \AA{ }emission is relatively constant in our sample. We furthermore find that 133 \AA{ }emission is typically more dominant in the lower energy, shorter duration events.

Our analysis shows trends in the peak intensities of EUV spectral lines with increasing emitted energy, however there are several spectral lines which do not show clear positive trends. For instance, though it increases in the most energetic events, the peak intensity of  284 \AA{ } emission does not show a consistent correlation with increased total radiation in the impulsive and intermediate flares. Additionally, as shown in Figure \ref {fig max_lc_intens},  the peak intensity of 304 \AA{ }\heii{ } emission ranges over an order of magnitude in the flares of our sample. For the both the impulsive (less energetic) and intermediate events there is no clear correlation with the total emitted energy. The most energetic flares in our sample always appear to always show large \heii{ }emission. However, given the dispersion present in the impulsive and intermediate events, the inclusion of more data is needed to concretely determine the behavior of the \heii{ }line in these cases.

\section{Discussion}

Correlations between the durations and energies in rise and decay phases suggest that the energy release process in flares is determined early in the flare evolution. Based on our results, several characteristics of the flare in its decay phase can be forecasted by observing the rise phase. We show that through constraining both the peak GOES flux and flare rise time, other physical parameters of a flare are determined; within our results, determination of two parameters is sufficient to roughly determine the flare's behavior. While this scaling applies to the single C8 class analyzed in this work, future work will investigate whether a similar scaling holds to additional flare classes. \cite{Veronig_2002} and \cite{Li_2012} showed that for a broad range of GOES classes, the time integrated flux of flare is proportional to the peak GOES flux times the duration. This result, consistent with our findings, suggests that several of the correlations reported within our study are applicable to additional flare classes. 

Constraining GOES class has proved a novel tool in the analysis of flare dynamics. In considering flares of multiple GOES classes, the correlations in energetic and temporal parameters within a single flare class are not immediately clear. {{With no constraint on peak SXR flux, these significant correlations within a single flare class are easily hidden within the statistical dispersion of a large sample set: e.g. Figure 3 of \cite{Veronig_2002}. }}

{{In addition to our investigation of the temporal and energetic aspects of a flare, the constraint of GOES class allows for a robust analysis of parameters such as temperature, emission measure, volume, and density. We have shown that within a single GOES class, the largest flares with the greatest amounts of emitted radiation are characteristically cooler, more voluminous, and less dense than less energetic events. As the temperatures of longer duration events are cooler, the emission measure of these events are correspondingly higher. The strength of these scaling laws as well as their applicability over the entire range of C8 flares will provide a useful foundation for future work, particularly regarding flare models and the development of scaling laws.}}

These empirically determined relations between flare parameters are straightforward and easily comparable with physical models. For instance, we have shown that a length diagnostic given for a single flaring loop by \cite{Reale_2007} compares surprisingly well with the observed data. It is interesting that this single loop model is able to reasonably diagnose loop lengths for such a range of flares: this model provides order of magnitude estimates for loop lengths of hot compact flares, events composed of a few closely bundled strands, as well as highly complex long duration events which incorporate vast numbers of coronal loops. These results are relevant in the context of stellar studies, where direct imaging of coronal plasma is not possible and indirect means of measurement are required to diagnose the spatial distribution of emitting coronal plasma. Assuming that similar results extend to different classes of flares, our findings suggest that this diagnostic, based on single loop hydrodynamic models of flares, is robust and provides sound estimates for flares which clearly involve several loop structures.

Scaling laws proposed by \cite{Serio_1991} relate the flare decay time with the temperature and length of the loop: $$t_b\propto \frac{L}{\sqrt{T_0}},$$  where $L$ is the loop length, $t_b$ is the decay time, $T_0$ is the peak flare temperature. The derivation of this scaling law considers both the conductive and radiative cooling times. \cite{Reale_1997} modified this law, including an empirically derived function $F$ to account for heating in the flare decay phase.

$$t_b=F \frac{L}{\sqrt{T_0}}$$

From these equations, it is clear that the duration of a flare increases with longer loop lengths and lower temperatures.
The diagnostics of Section \ref{subsec TEMP} show that the longer duration flares in our sample have characteristically lower temperatures and densities, as well as longer loop lengths. Accordingly, we expect these flares to have longer characteristic cooling times. Figure \ref{fig lc_peaks} shows that EVE observations are consistent with this analysis of cooling times: cooler flares with longer loop lengths are drawn out over extended periods of time.
 
The temporal evolution of parameters such as temperature and emission measure are also related to the flare cooling rate. Prolonged cooling times increase the intervals between the various stages of flare evolution. Therefore, in the long duration events, we expect an increased interval between the peak flare temperature and, for example, the peak SXR flux. In these flares, peak emission measure should occur relatively later in time: enhancements in coronal emission measure arise from the conductive heating of flare footpoints, resulting in the evaporation of chromospheric plasma. This process is prolonged by the longer cooling times. 


Previously, \cite{Aschwanden_2008} empirically determined scaling laws using statistical observations of both solar and stellar flares: 
  $$t_{tot}\propto T_p^{0.91 \pm 0.05}$$ 
  $${L}\propto T_p^{0.91 \pm 0.04},$$ 
where $T_p$ is the temperature at the peak flare time, $t_{tot}$, is the total duration and $L$ is the flare loop length. These power laws, which show positive relationships between temperature, duration, and loop length, represent global flare behavior without regard to GOES class. These power laws, derived from both extremely large and small events, observed on both the sun as well as other stars, show positive correlations between flare parameters such as loop length and temperature. However, our results show that within a single GOES class, these flare parameters are negatively correlated. {{These results show that global correlations are not applicable to constrained subsets of flares, such as the C8 flare class.

}}

 \section{Conclusion}

This study uses EVE and AIA data to analyze relatively isolated solar flares  within a small range around the C8 GOES class (events with peak flux around $8 \times 10^{-6}$ \wm). Through constraining our statistical analysis to flares of a single GOES class, we show the diversity of events which produce the same peak SXR flux. This work explicitly demonstrates how greatly duration, emitted radiation, temperature, volume, and density may vary within a single GOES flare class. In illustrating the range of events occurring within the C8 GOES class, we uncover significant relationships existing between various physical parameters. 

SDO and GOES observations of a sample of 17 C8 flares show strong relationships in the timings and energetics of the rise and the decay phases. We find that the duration of the rise phase is directly proportional to the decay phase time scale. Additionally, we show strong scaling between radiated SXR energy and duration. The amount of energy emitted by a C8 class flare is directly proportional to its duration. This proportionality further extends to the individual rise and decay phases: e.g. the energy emitted in the rise phase is directly proportional to the duration of the rise phase. We furthermore show a proportionality between the energy radiated in the rise and decay phases: this linear scaling is observed in both broadband EUV and SXR emission.

We have defined the rise and decay times using the full width quarter max (FWQM) values of the GOES 1-8 \AA{ }lightcurve; this selection is an arbitrary distinction. The particular values used to define the flare start and end times can significantly affect both the the durations and radiated energies reported for a flare. However, we have explored alternate definitions for time periods and verified the linear scaling between rise and total times of our sample of flares. We have furthermore verified that the SXR and EUV emitted energies during the rise and total duration of the flare scale linearly independent of the time definitions used.

It is especially interesting to find a highly correlated energy partitioning between the rise and decay phase in both the SXR emission (Log T $> 6.7$) as well as the the EUV band (sensitive to a temperature range of Log T $<$ 5 to Log T $>$ 7). EVE observations of our sample flares present a very broad range of spectral signatures. For instance, 304 \AA{ }emission from  \heii{ }is in some cases a dominant contributor to the emitted energy in the EVE band, but its peak emission is highly uncorrelated with the total energy of the flare and ranges by more than an order of magnitude for very similar flare energies (Figure \ref{fig sample_eve_lcs}). Analogously, the peak emission in the cooler (2-4MK) lines is very different from flare to flare and presents a large scatter. It is notable that the integrated EUV behavior is fairly predictable while the behavior of individual lines varies drastically.

In considering a variety of thermal aspects of these events, we find substantial correlations between numerous parameters. Our work demonstrates that within the C8 GOES class, flares with longer durations, which are generally more voluminous and have longer loop lengths, emit more total radiation. In turn, these events have lower temperatures and densities. Shorter duration flares, which are smaller in geometric size and emit the least radiation, are the hottest and densest events. Within this constrained GOES class, these relationships have a high certainty and limited statistical dispersion.

The time evolution of temperature and emission measure govern flare emission such as EUV spectral radiation. Through analyzing relationships between various thermal parameters, we explain observed variations in the spectral lines of these C8 events. We show that long duration energetic events tend to have brighter and longer enduring spectral emission. These events cool down over longer periods of time. Furthermore, these events tend to have prominent emission from spectral lines formed at temperatures less than Log T =7. The hottest events, which are also the fastest evolving, tend to be dominated by spectral lines formed at higher temperatures. Our investigation into relationships between the thermal properties of flares has allowed for a deeper understanding of their spectral emission.

The statistics of our study would be greatly improved by a significantly larger sample. Our study is impacted by the inclusion of the large filament eruption on 2012 August 31. Though this event is unique to our sample, it cannot be explicitly thought of as a statistical outlier. The flare is consistent with trends present in the rest of the data. To fully explore and quantify the relationships presented in this work, a larger statistical sample is needed; ideally a larger sample would include more of these large scale events. We have only included events which occurred after SDO data became available; however, it would be possible to expand the portions of this study which rely solely on GOES data, to include a much larger number of events ranging over solar cycles.

\begin{acknowledgements}

The authors would like to thank Ryan Milligan, John Raymond, and Phil Chamberlin for input which has greatly improved this paper. TB, PT, KR, were supported by contract SP02H1701R from Lockheed-Martin to Smithsonian Astrophysical Observatory. TB was partially supported under the NSF REU solar physics program at Harvard Smithsonian Center for Astrophysics, grant number ATM-0851866. PT also acknowledges support by NASA contract NNM07AB07C to the Smithsonian Astrophysical Observatory.
\end{acknowledgements}


\begin{thebibliography}{36}
\expandafter\ifx\csname natexlab\endcsname\relax\def\natexlab#1{#1}\fi

\bibitem[{{Aschwanden} {et~al.}(2008){Aschwanden}, {Stern}, \&
  {G{\"u}del}}]{Aschwanden_2008}
{Aschwanden}, M.~J., {Stern}, R.~A., \& {G{\"u}del}, M. 2008, \apj, 672, 659

\bibitem[{{Chamberlin} {et~al.}(2012){Chamberlin}, {Milligan}, \&
  {Woods}}]{Chamberlin_2012}
{Chamberlin}, P.~C., {Milligan}, R.~O., \& {Woods}, T.~N. 2012, \solphys, 279,
  23

\bibitem[{{Dere} {et~al.}(1997){Dere}, {Landi}, {Mason}, {Monsignori Fossi}, \&
  {Young}}]{Dere_1997_CHIANTI}
{Dere}, K.~P., {Landi}, E., {Mason}, H.~E., {Monsignori Fossi}, B.~C., \&
  {Young}, P.~R. 1997, \aaps, 125, 149

\bibitem[{{Dere} {et~al.}(2009){Dere}, {Landi}, {Young}, {Del Zanna},
  {Landini}, \& {Mason}}]{Dere_CHIANTI_Version}
{Dere}, K.~P., {Landi}, E., {Young}, P.~R., {Del Zanna}, G., {Landini}, M., \&
  {Mason}, H.~E. 2009, \aap, 498, 915

\bibitem[{{Donnelly}(1976)}]{Donnelly_1976}
{Donnelly}, R.~F. 1976, \jgr, 81, 4745

\bibitem[{{Drake}(1971)}]{Drake_1971}
{Drake}, J.~F. 1971, \solphys, 16, 152

\bibitem[{{Emslie} {et~al.}(2005){Emslie}, {Dennis}, {Holman}, \&
  {Hudson}}]{Emslie_2005}
{Emslie}, A.~G., {Dennis}, B.~R., {Holman}, G.~D., \& {Hudson}, H.~S. 2005,
  Journal of Geophysical Research (Space Physics), 110, 11103

\bibitem[{{Emslie} {et~al.}(2012){Emslie}, {Dennis}, {Shih}, {Chamberlin},
  {Mewaldt}, {Moore}, {Share}, {Vourlidas}, \& {Welsch}}]{Emslie_2012}
{Emslie}, A.~G., {Dennis}, B.~R., {Shih}, A.~Y., {Chamberlin}, P.~C.,
  {Mewaldt}, R.~A., {Moore}, C.~S., {Share}, G.~H., {Vourlidas}, A., \&
  {Welsch}, B.~T. 2012, \apj, 759, 71

\bibitem[{{Feldman} {et~al.}(1996){Feldman}, {Doschek}, {Behring}, \&
  {Phillips}}]{Feldman_1996_temperature}
{Feldman}, U., {Doschek}, G.~A., {Behring}, W.~E., \& {Phillips}, K.~J.~H.
  1996, \apj, 460, 1034

\bibitem[{{Feldman} {et~al.}(1997){Feldman}, {Doschek}, \&
  {Klimchuk}}]{Feldman_1997}
{Feldman}, U., {Doschek}, G.~A., \& {Klimchuk}, J.~A. 1997, \apj, 474, 511

\bibitem[{{Garcia}(1994)}]{Garcia_1994}
{Garcia}, H.~A. 1994, \solphys, 154, 275

\bibitem[{{Hock} {et~al.}(2012){Hock}, {Woods}, {Klimchuk}, {Eparvier}, \&
  {Jones}}]{Hock_2012}
{Hock}, R.~A., {Woods}, T.~N., {Klimchuk}, J.~A., {Eparvier}, F.~G., \&
  {Jones}, A.~R. 2012, ArXiv e-prints

\bibitem[{{Lee} {et~al.}(1995){Lee}, {Petrosian}, \& {McTiernan}}]{Lee_1995}
{Lee}, T.~T., {Petrosian}, V., \& {McTiernan}, J.~M. 1995, \apj, 448, 915

\bibitem[{{Lemen} {et~al.}(2012){Lemen}, {Title}, {Akin}, {Boerner}, {Chou},
  {Drake}, {Duncan}, {Edwards}, {Friedlaender}, {Heyman}, {Hurlburt}, {Katz},
  {Kushner}, {Levay}, {Lindgren}, {Mathur}, {McFeaters}, {Mitchell}, {Rehse},
  {Schrijver}, {Springer}, {Stern}, {Tarbell}, {Wuelser}, {Wolfson}, {Yanari},
  {Bookbinder}, {Cheimets}, {Caldwell}, {Deluca}, {Gates}, {Golub}, {Park},
  {Podgorski}, {Bush}, {Scherrer}, {Gummin}, {Smith}, {Auker}, {Jerram},
  {Pool}, {Soufli}, {Windt}, {Beardsley}, {Clapp}, {Lang}, \&
  {Waltham}}]{Lemen_AIA}
{Lemen}, J.~R., {Title}, A.~M., {Akin}, D.~J., {Boerner}, P.~F., {Chou}, C.,
  {Drake}, J.~F., {Duncan}, D.~W., {Edwards}, C.~G., {Friedlaender}, F.~M.,
  {Heyman}, G.~F., {Hurlburt}, N.~E., {Katz}, N.~L., {Kushner}, G.~D., {Levay},
  M., {Lindgren}, R.~W., {Mathur}, D.~P., {McFeaters}, E.~L., {Mitchell}, S.,
  {Rehse}, R.~A., {Schrijver}, C.~J., {Springer}, L.~A., {Stern}, R.~A.,
  {Tarbell}, T.~D., {Wuelser}, J.-P., {Wolfson}, C.~J., {Yanari}, C.,
  {Bookbinder}, J.~A., {Cheimets}, P.~N., {Caldwell}, D., {Deluca}, E.~E.,
  {Gates}, R., {Golub}, L., {Park}, S., {Podgorski}, W.~A., {Bush}, R.~I.,
  {Scherrer}, P.~H., {Gummin}, M.~A., {Smith}, P., {Auker}, G., {Jerram}, P.,
  {Pool}, P., {Soufli}, R., {Windt}, D.~L., {Beardsley}, S., {Clapp}, M.,
  {Lang}, J., \& {Waltham}, N. 2012, \solphys, 275, 17

\bibitem[{{Li} {et~al.}(2012){Li}, {Gan}, \& {Feng}}]{Li_2012}
{Li}, Y.~P., {Gan}, W.~Q., \& {Feng}, L. 2012, \apj, 747, 133

\bibitem[{{Mart{\'{\i}}nez-Sykora} {et~al.}(2011){Mart{\'{\i}}nez-Sykora}, {De
  Pontieu}, {Testa}, \& {Hansteen}}]{Martinez_Sykora_2011}
{Mart{\'{\i}}nez-Sykora}, J., {De Pontieu}, B., {Testa}, P., \& {Hansteen}, V.
  2011, \apj, 743, 23

\bibitem[{{Milligan}(2011)}]{Milligan_2011}
{Milligan}, R.~O. 2011, \apj, 740, 70

\bibitem[{{Milligan} {et~al.}(2012{\natexlab{a}}){Milligan}, {Chamberlin},
  {Hudson}, \& {Woods}}]{Milligan_2012_continuum}
{Milligan}, R.~O., {Chamberlin}, P.~C., {Hudson}, H.~S., \& {Woods}, T.~N.
  2012{\natexlab{a}}, \apjl, 748, L14

\bibitem[{{Milligan} {et~al.}(2012{\natexlab{b}}){Milligan}, {Kennedy},
  {Mathioudakis}, \& {Keenan}}]{Milligan_2012_density}
{Milligan}, R.~O., {Kennedy}, M.~B., {Mathioudakis}, M., \& {Keenan}, F.~P.
  2012{\natexlab{b}}, \apjl, 755, L16

\bibitem[{{O'Dwyer} {et~al.}(2010){O'Dwyer}, {Del Zanna}, {Mason}, {Weber}, \&
  {Tripathi}}]{O'Dwyer_2010}
{O'Dwyer}, B., {Del Zanna}, G., {Mason}, H.~E., {Weber}, M.~A., \& {Tripathi},
  D. 2010, \aap, 521, A21

\bibitem[{{Pearce} \& {Harrison}(1988)}]{Pearce_1988}
{Pearce}, G. \& {Harrison}, R.~A. 1988, \aap, 206, 121

\bibitem[{{Pearce} {et~al.}(1993){Pearce}, {Rowe}, \& {Yeung}}]{Pearce_1993}
{Pearce}, G., {Rowe}, A.~K., \& {Yeung}, J. 1993, \apss, 208, 99

\bibitem[{{Reale}(2007)}]{Reale_2007}
{Reale}, F. 2007, \aap, 471, 271

\bibitem[{{Reale} {et~al.}(1997){Reale}, {Betta}, {Peres}, {Serio}, \&
  {McTiernan}}]{Reale_1997}
{Reale}, F., {Betta}, R., {Peres}, G., {Serio}, S., \& {McTiernan}, J. 1997,
  \aap, 325, 782

\bibitem[{{Reeves} \& {Golub}(2011)}]{Reeves_Golub_2011}
{Reeves}, K.~K. \& {Golub}, L. 2011, \apjl, 727, L52

\bibitem[{{Reeves} \& {Warren}(2002)}]{Reeves_Warren_2002}
{Reeves}, K.~K. \& {Warren}, H.~P. 2002, \apj, 578, 590

\bibitem[{{Reeves} {et~al.}(2007){Reeves}, {Warren}, \& {Forbes}}]{Reeves_2007}
{Reeves}, K.~K., {Warren}, H.~P., \& {Forbes}, T.~G. 2007, \apj, 668, 1210

\bibitem[{{Ryan} {et~al.}(2012){Ryan}, {Milligan}, {Gallagher}, {Dennis},
  {Tolbert}, {Schwartz}, \& {Young}}]{Ryan_2012_TEBBS}
{Ryan}, D.~F., {Milligan}, R.~O., {Gallagher}, P.~T., {Dennis}, B.~R.,
  {Tolbert}, A.~K., {Schwartz}, R.~A., \& {Young}, C.~A. 2012, \apjs, 202, 11

\bibitem[{{Serio} {et~al.}(1991){Serio}, {Reale}, {Jakimiec}, {Sylwester}, \&
  {Sylwester}}]{Serio_1991}
{Serio}, S., {Reale}, F., {Jakimiec}, J., {Sylwester}, B., \& {Sylwester}, J.
  1991, \aap, 241, 197

\bibitem[{{Shimizu}(1995)}]{Shimizu_1995}
{Shimizu}, T. 1995, \pasj, 47, 251

\bibitem[{{Veronig} {et~al.}(2002){Veronig}, {Temmer}, {Hanslmeier}, {Otruba},
  \& {Messerotti}}]{Veronig_2002}
{Veronig}, A., {Temmer}, M., {Hanslmeier}, A., {Otruba}, W., \& {Messerotti},
  M. 2002, \aap, 382, 1070

\bibitem[{{Warren}(2006)}]{Warren_2006}
{Warren}, H.~P. 2006, \apj, 637, 522

\bibitem[{{White} {et~al.}(2005){White}, {Thomas}, \& {Schwartz}}]{White_2005}
{White}, S.~M., {Thomas}, R.~J., \& {Schwartz}, R.~A. 2005, \solphys, 227, 231

\bibitem[{{Woods} {et~al.}(2012){Woods}, {Eparvier}, {Hock}, {Jones},
  {Woodraska}, {Judge}, {Didkovsky}, {Lean}, {Mariska}, {Warren}, {McMullin},
  {Chamberlin}, {Berthiaume}, {Bailey}, {Fuller-Rowell}, {Sojka}, {Tobiska}, \&
  {Viereck}}]{Woods_2011_EVE}
{Woods}, T.~N., {Eparvier}, F.~G., {Hock}, R., {Jones}, A.~R., {Woodraska}, D.,
  {Judge}, D., {Didkovsky}, L., {Lean}, J., {Mariska}, J., {Warren}, H.,
  {McMullin}, D., {Chamberlin}, P., {Berthiaume}, G., {Bailey}, S.,
  {Fuller-Rowell}, T., {Sojka}, J., {Tobiska}, W.~K., \& {Viereck}, R. 2012,
  \solphys, 275, 115

\bibitem[{{Woods} {et~al.}(2011){Woods}, {Hock}, {Eparvier}, {Jones},
  {Chamberlin}, {Klimchuk}, {Didkovsky}, {Judge}, {Mariska}, {Warren},
  {Schrijver}, {Webb}, {Bailey}, \& {Tobiska}}]{Woods_2011_latephase}
{Woods}, T.~N., {Hock}, R., {Eparvier}, F., {Jones}, A.~R., {Chamberlin},
  P.~C., {Klimchuk}, J.~A., {Didkovsky}, L., {Judge}, D., {Mariska}, J.,
  {Warren}, H., {Schrijver}, C.~J., {Webb}, D.~F., {Bailey}, S., \& {Tobiska},
  W.~K. 2011, \apj, 739, 59

\bibitem[{{Zhang} {et~al.}(2012){Zhang}, {Cai}, {Ercha}, {Hao}, \&
  {Xiao}}]{Zhang_2012}
{Zhang}, D.~H., {Cai}, L., {Ercha}, A., {Hao}, Y.~Q., \& {Xiao}, Z. 2012,
  \solphys, 280, 183

\end{thebibliography}
\end{document}